\documentclass[sigconf]{acmart}

\AtBeginDocument{%
  \providecommand\BibTeX{{%
    \normalfont B\kern-0.5em{\scshape i\kern-0.25em b}\kern-0.8em\TeX}}}

\copyrightyear{2021}
\acmYear{2021}
\setcopyright{acmcopyright}
\acmConference[MICRO '21]{MICRO-54: 54th Annual IEEE/ACM International Symposium on Microarchitecture}{October 18--22, 2021}{Virtual Event, Greece}
\acmBooktitle{MICRO-54: 54th Annual IEEE/ACM International Symposium on Microarchitecture (MICRO '21), October 18--22, 2021, Virtual Event, Greece}
\acmPrice{15.00}
\acmDOI{10.1145/3466752.3480120}
\acmISBN{978-1-4503-8557-2/21/10}

\begin{document}

\title{Shift-BNN: Highly-Efficient Probabilistic Bayesian Neural Network Training via Memory-Friendly Pattern Retrieving}

\author{Qiyu Wan}
\affiliation{%
  \institution{University of Houston}
  \city{Houston}
  \state{Texas}
  \country{USA}
}
\email{qwan@uh.edu}

\author{Haojun Xia}
\affiliation{%
  \institution{University of Sydney}
  \city{Sydney}
  \country{Australia}}
\email{xhjustc@gmail.com}

\author{Xingyao Zhang}
\affiliation{%
  \institution{University of Washington}
  \city{Seattle}
  \state{Washington}
  \country{USA}
}
\email{xingyaoz@cs.washington.edu}

\author{Lening Wang}
\affiliation{%
 \institution{University of Houston}
 \city{Houston}
 \state{Texas}
 \country{USA}}
\email{lwang56@uh.edu}

\author{Shuaiwen Leon Song}
\affiliation{%
  \institution{University of Sydney}
  \city{Sydney}
  \country{USA}}
\email{leonangel991@gmail.com}

\author{Xin Fu}
\affiliation{%
  \institution{University of Houston}
  \city{Houston}
  \state{Texas}
  \country{USA}}
\email{xfu8@central.uh.edu}

\renewcommand{\shortauthors}{Qiyu Wan, et al.}

\begin{abstract}
Bayesian Neural Networks (\textit{BNNs}) that possess a property of uncertainty estimation have been increasingly adopted in a wide range of safety-critical AI applications which demand reliable and robust decision making, e.g., self-driving, rescue robots, medical image diagnosis. The training procedure of a probabilistic BNN model involves training an ensemble of sampled DNN models, which induces orders of magnitude larger volume of data movement than training a single DNN model. In this paper, we reveal that the root cause for BNN training inefficiency originates from the massive off-chip data transfer by Gaussian Random Variables (\textit{GRVs}). To tackle this challenge, we propose a novel design that eliminates all the off-chip data transfer by GRVs through the reversed shifting of Linear Feedback Shift Registers (\textit{LFSRs}) without incurring any training accuracy loss. To efficiently support our LFSR reversion strategy at the hardware level, we explore the design space of the current DNN accelerators and identify the optimal computation mapping scheme to best accommodate our strategy. By leveraging this finding, we design and prototype the first highly efficient BNN training accelerator, named \textit{Shift-BNN}, that is low-cost and scalable.  Extensive evaluation on five representative BNN models demonstrates that Shift-BNN achieves an average of 4.9$\times$ (up to 10.8$\times$) boost in energy efficiency and 1.6$\times$ (up to 2.8$\times$) speedup over the baseline DNN training accelerator.
\end{abstract}






\maketitle

\section{Introduction}
Deep learning based AI technologies, such as deep convolutional neural networks (DNNs), have recently achieved tremendous success in numerous application domains, such as object detection, image classification, etc \cite{szegedy2013deep,bojarski2016end, farooq2017deep,cheron2015p,simonyan2014two}. However, DNN models are known to be prone to over-fitting due to insufficient training data in the real world, which can lead to wrong predictions when the model is deployed in unfamiliar environments. With the increasing adaptation of safety-critical AI applications (e.g., healthcare and self-driving), wrong predictions can result in catastrophic incidents. For example, several accidents have been recently reported regarding poor safety-critical AI designs \cite{tesla2018,car1}, e.g., in 2020 an autopilot car crashed into a white truck because the sensor failed to distinguish the truck from the bright sky \cite{car1}. Therefore, enhancing the reliability and robustness of deep learning has become an urgent demand from AI practitioners. 

As one of the most popular probabilistic machine learning tools, Bayesian Neural Networks (BNNs) have been increasingly employed in a wide range of real-world AI applications which require reliable and robust decision making such as self-driving, rescue robots, disease diagnosis, scene understanding, and so on \cite{amini2018spatial,wulfmeier2018machine,leibig2017leveraging,kendall2015bayesian}. BNNs have also emerged as a promising solution in today’s data center services for improving product experiences (e.g., Instagram and Youtube), infrastructure, and aiding cutting-edge research \cite{Fb}. Different from the traditional DNNs which require massive training data, BNN models can more easily learn from small datasets and are more robust to over-fitting issues \cite{blundell2015weight}. Furthermore, BNNs are capable of providing valuable uncertainty information for users to better interpret the situation without making over-confident decisions \cite{gal2016dropout, amodei2016concrete,cheung2011bayesian}. Generally, a BNN model can be viewed as a probabilistic model where each model parameter, i.e., weight, is a probability distribution. Training a BNN essentially calculates the probability distribution of weights, which requires integrating on infinite number of neural networks. This is often intractable. To tackle this, recent efforts \cite{blundell2015weight,graves2011practical,shridhar2019comprehensive} leverage Gaussian distributions to approximate the target weight distributions via \textit{weight sampling} to identify the mean and standard deviation of each weight.

Training DNN models on current hardware devices has long been considered as a slow and energy-consuming task \cite{goel2020survey,venkataramani2017scaledeep,das2016distributed}. Compared with the traditional DNN training, BNN training inefficiency is further exacerbated by the requirement of \textit{training an ensemble of sampled DNN models} to ensure robustness. In consequence, we have observed that the total \textit{data movement} during a BNN training procedure can be orders of magnitude larger than training one single DNN model. Moreover, as the existing DNN training optimization techniques \cite{zheng2020echo,rhu2016vdnn,zhang2019eager,yang2020procrustes,wang2018superneurons} are oblivious to the unique sampling process of the probabilistic BNN models, they lack the capabilities to efficiently and effectively deal with the excessive data movement induced by the memory-intensive BNN training, resulting in poor energy efficiency and long training latency.

\begin{figure*}
\centering
	\includegraphics[scale=0.67]{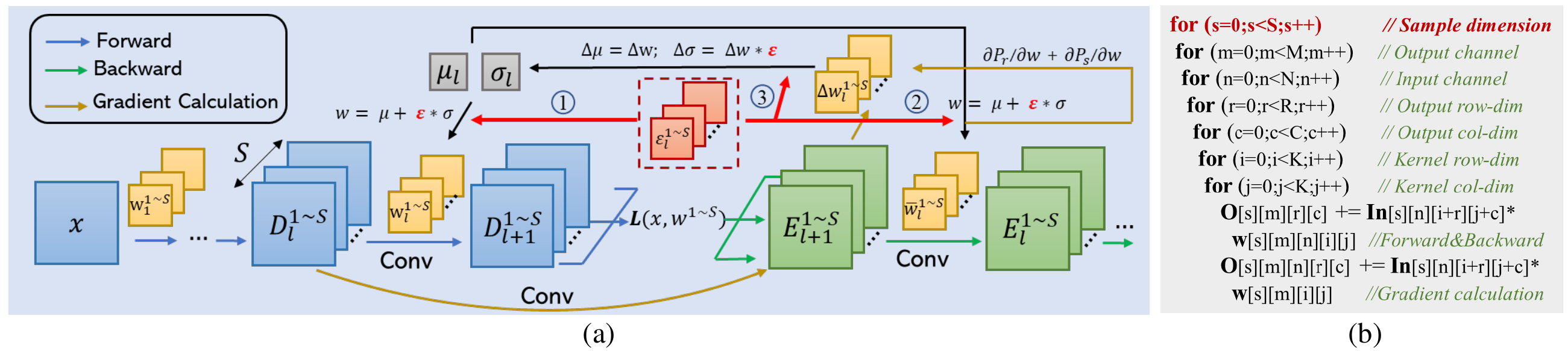}
	\vspace*{-2.5mm}
	\caption{(a) Computation flow of BNN training. (b) 7-dimension for-loop for the convolutional layer in BNN.}
	\label{fig:bg}
	\vspace*{-2.5mm}
\end{figure*}

In this paper, we first conduct a comprehensive characterization of the state-of-the-art BNN training on current DNN accelerators and analyze its inefficiency. 
By carefully breaking down the memory activities in each BNN layer, we observe that the dominant factor that induces BNN training inefficiency is the massive data movement from Gaussian random variables (GRVs). These variables are generated during forward propagation for weight sampling and sent to off-chip memory for later reuse during backward propagation. They contribute the major portion of the total off-chip memory accesses for BNN training (e.g., up to 71\%). To tackle this challenge, we propose a novel design that is capable of \textit{eliminating all the off-chip memory accesses by GRVs without incurring any training accuracy loss.} Our design is based on a key observation that the software-level ``forth-back" training procedure shares great similarity with the classic hardware-level reversed shifting of the Linear Feedback Shift Registers (LFSRs) which are used in modern BNNs to generate the GRVs \cite{cai2018vibnn}. By leveraging the reversible property of LFSR, we build a highly efficient memory-friendly design based on LFSR reversed shifting, which can accurately retrieve all the GRVs (i.e., bit patterns generated in forward propagation) locally during backpropagation without ever storing them during the forward propagation. Furthermore, to investigate the compatibility of our LFSR reversion strategy on real hardware, we qualitatively study the design possibilities by directly integrating our strategy to the existing DNN accelerators that adopt various computation mapping schemes, and eventually identify the optimal mapping to support our BNN training design. Based on this knowledge, we design and prototype the first highly efficient hardware accelerator for BNN training, named \textit{Shift-BNN}, that takes advantage of drastically reduced data movement enabled by our LFSR reversion strategy. This study makes the following contributions:
\vspace*{-0.5mm}
\begin{itemize}
    \item We characterize modern BNN training on the state-of-the-art DNN accelerators and reveal that the root cause for its training inefficiency originates from the massive data transfer induced by GRVs;
    
    \item We propose a novel design that eliminates all the off-chip data transfer related to GRVs through local LFSR reversed shifting without affecting the training accuracy;
    
    \item We present the potential hardware-level challenges when directly applying our design to BNN training and significantly mitigate these issues via a sophisticated and qualitative design space exploration;
    
    \item We design and prototype the first highly-efficient BNN training accelerator that is low-cost and scalable, well supported by a hybrid dataflow;
    
    \item Extensive evaluation on five representative BNN models demonstrates that Shift-BNN achieves an average of 4.9$\times$ (up to 10.8$\times$) improvement in energy efficiency and 1.6$\times$ (up to 2.8$\times$) speedup over the baseline accelerator. Shift-BNN also scales well to larger BNN model sample sizes.
\end{itemize}

\section{Background}

\subsection{Training BNNs with Variational Inference}
% 1. Briefly introduce the principle of BNN training.
A Bayesian neural network (BNN) can be viewed as a probabilistic model in which each model parameter,e.g., weight, is a probability distribution. One of the most popular method for training BNN models is known as \textit{Variational Inference} \cite{blei2017variational,hoffman2013stochastic,zhang2018advances} , which finds a probability distribution $q(\mathbf{w}|\theta) \in \mathcal{Q}$ to approximate the target weight distribution ($\mathcal{Q}$ is a common distribution family). Searching for $q(\mathbf{w}|\theta)$ is an optimization problem that aims to minimize the loss function with respect to $\theta$ (Eq.\ref{eq2}).

\vspace{-4.5mm}

\begin{equation}\small
\begin{aligned}
\mathcal{L}(\mathbf{w}, \theta) = \sum_{i=1}^{S}\log q(\mathbf{w^{(i)}}|\theta) - \log P(\mathbf{w^{(i)}}) - \log P(\mathbf{y}|\mathbf{x, w^{(i)}})
\label{eq2}
\end{aligned}   
\end{equation}

\vspace{-2mm}
In Eq.\ref{eq2}, $\mathbf{w^{(i)}}$ denotes the $i$th sample of weights drawn from the approximation distribution $q(\mathbf{w}|\theta)$. Typically, $q(\mathbf{w}|\theta)$ is assumed to be a Gaussian distribution $q(\mathbf{w}|(\mu, \sigma))$ where $\mu$ and $\sigma$ are the mean and standard deviation of the Gaussian distribution, respectively. Each sample of weight $\mathbf{w^{(i)}}$ can be obtained by using $\mathbf{w^{(i)}} = \mu + \epsilon^\mathbf{(i)}  \circ  \sigma$, where $\epsilon^\mathbf{(i)}$ denotes the $i$th random variable drawn from unit Gaussian distribution $\mathcal{N}(0, \mathbf{I})$ and $\circ$ represent point-wise multiplication.
$\log q(\mathbf{w^{(i)}}|\theta)$, $\log P(\mathbf{w^{(i)}})$ and $\log P(\mathbf{y}|\mathbf{x, w^{(i)}})$ are defined as posterior $P_s$ , prior $P_r$ and log-likelihood, respectively. In summary, the model parameters $\mu$ and $\sigma$ can be learned progressively by repeating the following steps (details are shown in Fig.\ref{fig:bg} (a)): 

\vspace*{-0.5mm}
\begin{itemize}
    \item [1.]
    Generate S $\epsilon$'s from $\mathcal{N}(0, \mathbf{I})$ for each weight;
    
    \item [2.] 
    Obtain S samples for each weight via {\small $\mathbf{w^{(i)}} = \mu +\epsilon^\mathbf{(i)}  \circ  \sigma$};
    
    \item [3.]
    Calculate the loss function $\mathcal{L}(\mathbf{w}, \theta)$, where $\theta = (\mu, \sigma)$;
    
    \item [4.]
    Calculate the gradients with respect to $\mu$ and $\sigma$;
    
    \item [5.]
    Update model parameters $\mu$ and $\sigma$.
\end{itemize}

\subsection{Computation Flow of BNN Training} \label{cf}
From an algorithmic perspective, Fig.\ref{fig:bg} (a) illustrates the computation flow of BNN training which consists of three main stages: Forward (FW), Backward (BW) and Gradient Calculation (GC).

\textbf{Forward (FW)} stage aims to calculate the loss of network function $\mathbf{f}$ given an input training example $\mathbf{x}$. For simplicity of discussion, we assume processing a minibatch with the size of 1. In each layer \textit{l}, for one input training example, Gaussian random variables $\epsilon_l$ are sampled S times to obtain S samples of weights $w_l^1 \sim w_l^S$, denoted as process \textcircled{\small{1}}. These weights are convolved with their corresponding input samples, i.e., $D_l^1 \sim D_l^S$, producing S samples of the output, which are then treated as the input for the next layer. For the first layer, all weight samples are convolved with the input $\mathbf{x}$. The outputs of the last layer are compared with the groundtruth to obtain the loss (error).

\textbf{Backward (BW)} stage propagates the network errors from the last layer to the first layer. In each layer \textit{l}, S samples of weight matrices are reconstructed using the original Gaussian random variables $\epsilon$ and model parameters $(\mu,\sigma)$, denoted as process \textcircled{\small{2}}. The reconstructed kernels are then rotated $180^{\circ}$ and convolved with the corresponding samples of errors to obtain the errors of the previous layer, i.e., $E_l^1 \sim E_l^S$.

\textbf{Gradient Calculation (GC)} stage updates the model parameters $\mu$ and $\sigma$ to minimize the training loss, which requires to calculate the gradients of the model parameters, $\Delta\mu$ and $\Delta\sigma$. The gradient of a sampled weight comes from prior $P_r$, posterior $P_s$ and likelihood $P(\mathbf{y}|\mathbf{x, w^{(i)}})$. The gradient of likelihood is generated by convolving the feature maps $D_l^1 \sim D_l^S$ with the errors $E_{l+1}^1 \sim E_{l+1}^S$. This part is the same as the normal DNN training. For the gradients of prior and posterior, they can be easily derived once the original weights are reconstructed because the computation for both prior and posterior requires no intermediate feature maps.
Finally, the S samples of the gradients are summed up and then multiplied with a small coefficient to produce the weight updates $\Delta w$. Based on the sampling rule $w=\mu+\epsilon*\sigma$, Gaussian random variables $\epsilon$ are used to calculate the final updates $\Delta\sigma$. This step corresponds to step \textcircled{\small{3}}.

Fig.\ref{fig:bg} (b) illustrates the detailed computation within a single BNN's convolutional layer. The key feature here is a sample dimension that adds on top of normal DNNs' 6-dimension convolution. Note that different samples execute independently without any data exchange.
\vspace*{-3mm}

\section{Challenges of BNN training} \label{challenge}

% Off-chip data transfer of random variable $\epsilon$ takes a major percentage. 
\textbf{Traditional DNN training.} DNN training has long been considered as a slow and energy harvesting task  \cite{goel2020survey,venkataramani2017scaledeep,das2016distributed}. On the surface, the massive energy consumption and high latency mainly come from millions of Multiply-accumulate operations (MACs) and intensive data movement between memory and processing elements (PEs). As the unit energy cost ({J/bit}) of off-chip memory accesses is orders of magnitude higher than that of MACs \cite{chen2016eyeriss,dally2011power,horowitz20141}, data movement usually poses greater challenges for energy-efficient DNN training \cite{wang2019e2}. Moreover, the ongoing development of low-precision training techniques \cite{gupta2015deep,wang2018training, fu2020fractrain} can potentially reduce the unit energy cost of MACs, but this could also result in a proportionally higher impact on the overall training's energy efficiency from the data movement.  

%in comparison data movement's impact on the overall training's energy efficiency appears to be more prominent. 
\begin{figure}[t]
\centering
\includegraphics[scale=0.59]{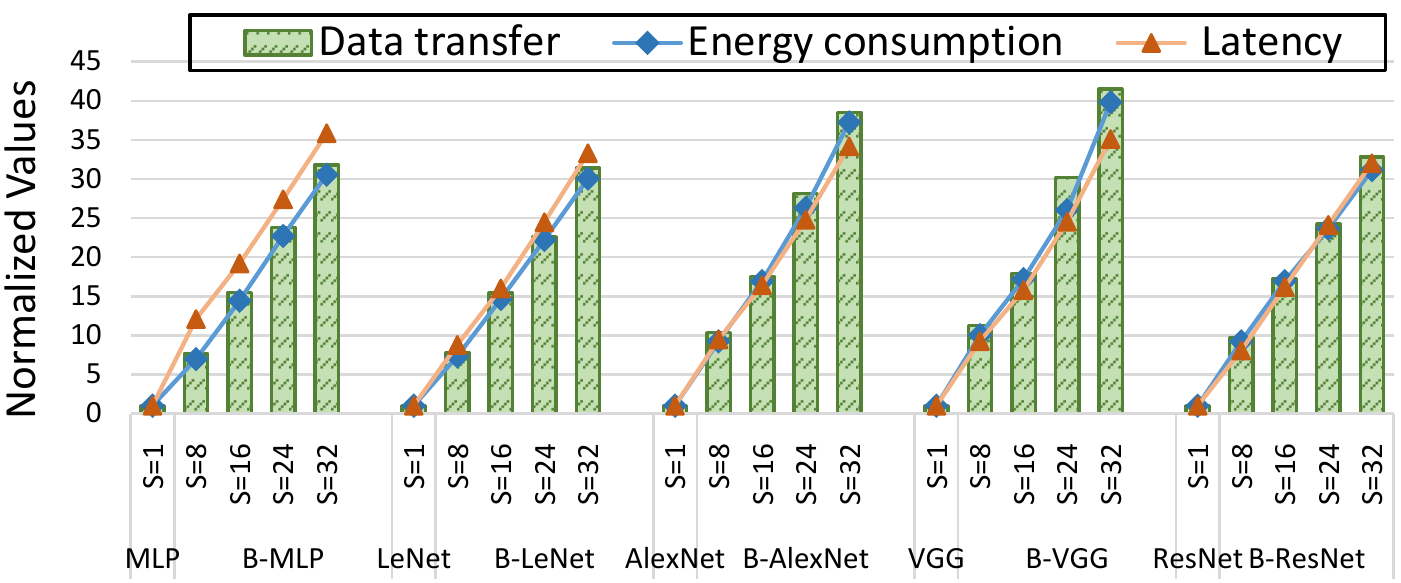}
\vspace*{-2.5mm}
\caption{Comparison between five BNN models and their corresponding baseline DNN models. }
\label{fig:motiv}
\vspace*{-5.5mm}
\end{figure}

\textbf{Current BNN training.} Compared to the traditional DNN training, BNN training inefficiency is further exacerbated by the requirement of training for \textit{an ensemble of sampled DNN models}, shown in Fig. \ref{fig:bg} (a). This is necessary because a sufficient number of training samples is essential for building a robust BNN model. But it could also incur an explosive amount of data movement during the training process. 

To further quantify this, we investigate the impact of number of samples on the overall BNN training efficiency. We implemented five types of widely-adopted BNN models representing a broad range of domains, as well as their corresponding DNN models. Note that BNN models are typically built upon their matching DNN models, e.g., Bayesian AlexNet or \textit{B-Alexnet} is based on AlexNet. For verification purposes, the training process is performed on a general Diannao-like DNN accelerator equipped with output stationary dataflow \cite{chen2014diannao}. Detailed experimental setup can be found in Section \ref{setup}. Three metrics are used for training evaluation, including data transfer, overall energy consumption, and training latency. The data transfer represents the amount of data that are read from and written to the off-chip memory. Due to the architectural heterogeneity of the five BNN models, each result is normalized to its corresponding baseline DNN model. 
Fig.\ref{fig:motiv} shows that a BNN model with only 8 samples would drastically increase the off-chip data transfer by an average of 9.1$\times$ compared with its corresponding DNN model. This number grows to 35.3$\times$ as the number of BNN training samples scales up to 32. Specifically, for B-VGG model with 16 samples (s=16), training each input example for one iteration would require 22.6GB data transfer from/to off-chip memory, which is 17.9$\times$ increment over the original VGG model. Since the off-chip memory access is often considered a high-cost operation, a large amount of data transfer during BNN training could produce massive energy consumption and potentially lead to performance degradation. For example, we observed that the overall energy consumption and training latency on 32 samples incur an average of 33.2$\times$ and 31.8$\times$ increment over those on the baseline DNN models, respectively. 

\begin{figure}[t]
\centering
\includegraphics[scale=0.6]{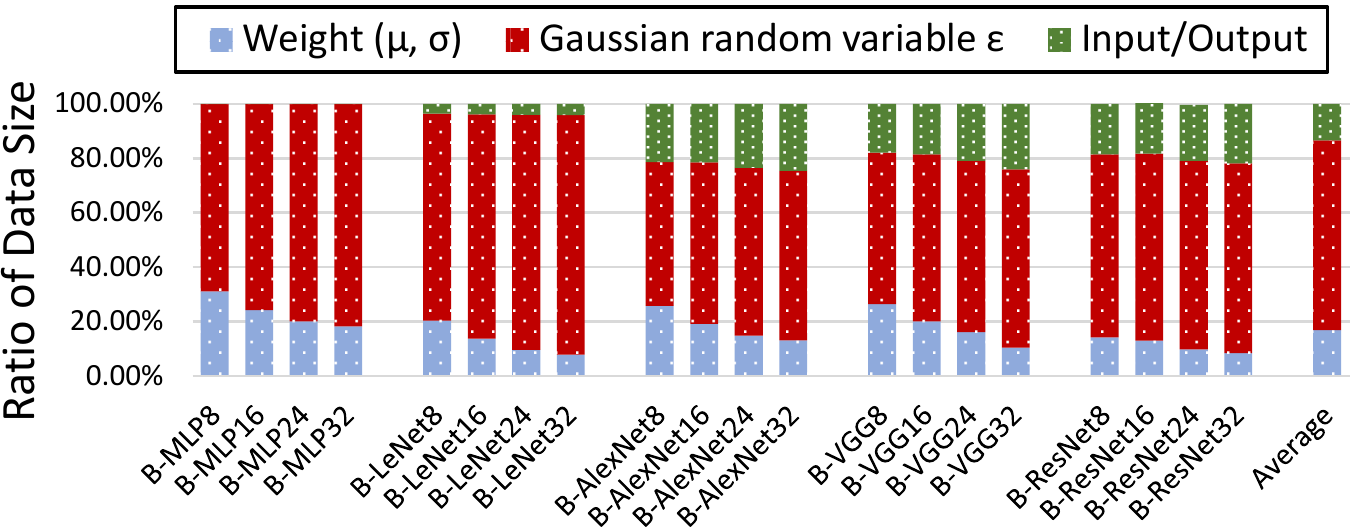}
\vspace*{-3.5mm}
\caption{The ratio breakdown of the total off-chip data transfer across different BNN models.}

\label{fig:data_size}
\vspace*{-5.5mm}
\end{figure}

Fig.\ref{fig:data_size} shows the breakdown of the total off-chip data transfer when the accelerator evaluates every input training example during one training iteration. It can be observed that Gaussian random variables $\epsilon$ takes up the major portion of the total data transfer (i.e., 71\% on average). Meanwhile, the weight parameters $(\mu, \sigma)$ and the input/output feature maps only contribute to 16\% and 12\% on average, respectively. There are several reasons behind such dominating presence of $\epsilon$. First, as a unique variable introduced by BNN execution, $\epsilon$ must be stored and reused in two different stages. As shown in Fig.\ref{fig:bg} (a) \textcircled{\small{1}}, during the forward stage, S samples of $\epsilon$ are generated from the local random number generators for each pair of $(\mu, \sigma)$ to obtain S samples of weights. After that, $\epsilon$s have to be stored into the off-chip memory due to its large data volume and reside there until the later weight reconstruction during the backward stage (\textcircled{\small{2}}) and the gradient of $\sigma$ computation during the gradient calculation (GC) stage (\textcircled{\small{3}}). Note that recent memory-centric approaches such as vDNN \cite{rhu2016vdnn}, Echo \cite{zheng2020echo} and SuperNeurons \cite{wang2018superneurons} reduce the memory accesses through smart recomputation in backpropagation via selected small intermediate data from forward propagation. However, since $\epsilon$s are a large amount of independent random numbers that cannot be recomputed, these works cannot help reduce intensive memory accesses in BNN training. Second, the size of $\epsilon$ is much larger than the weight parameters $(\mu, \sigma)$ and the intermediate feature maps/errors. Since one pair of weight parameters $(\mu, \sigma)$ requires S samples of $\epsilon$ for weight sampling, the total size of $\epsilon$ can be $S/2$ times of the weight parameters. And for the current BNN models, the size of weights (i.e., half of $(\mu, \sigma)$) is still much larger than the size of feature maps. For instance, on average the size of weights is 122$\times$ of the size of feature maps/errors across five BNN models. Therefore, although input/output feature maps also consist of S samples, the total transferred intermediate data size is still much less than that from $\epsilon$.

In summary, the long reuse distance of a large amount of Gaussian random variables $\epsilon$ across different training stages is the key problem that causes a huge amount of off-chip memory accesses (the transferred amount of $\epsilon$s grows linearly with the sample size). This further leads to massive energy consumption and potential performance degradation during BNN training. Besides the existing DNN accelerator, such a challenge is also observed on conventional CPU/GPU platforms as the cross-stage memory access of $\epsilon$ is inevitable in the BNN training algorithm. Therefore, a special solution is needed.

\section{Key Design Insights of Shift-BNN} \label{insight}
To overcome these challenges brought by the excessive data movement for the Gaussian random variables $\epsilon$s (or GRVs), we propose a novel design that is able to \textit{eliminate all the memory accesses related to $\epsilon$ without training accuracy loss}. We made a key observation that the nature of software-level``forth-back" training procedure shares similarity with the classic hardware-level reversed shifting of Linear Feedback Shift Register (LFSR) which is used in BNNs to generate the Gaussian random variables $\epsilon$ \cite{cai2018vibnn}. Specifically, we can potentially retrieve all the $\epsilon$s locally during the Backward stage through \textit{shifting the LFSRs backward}, instead of storing them during the Forward stage. In the following subsections, we will first introduce the principles of LFSR function, and then illustrate how to use LFSR reversed shifting to retrieve Gaussian random variables $\epsilon$s. Finally, we showcase a detailed example to demonstrate the feasibility of our strategy while also exposing some potential hardware-level issues when directly applying it to BNN training. 

\subsection{Generating GRVs via LFSR Shifting}
\begin{figure}[t]
\centering
\includegraphics[scale=0.49]{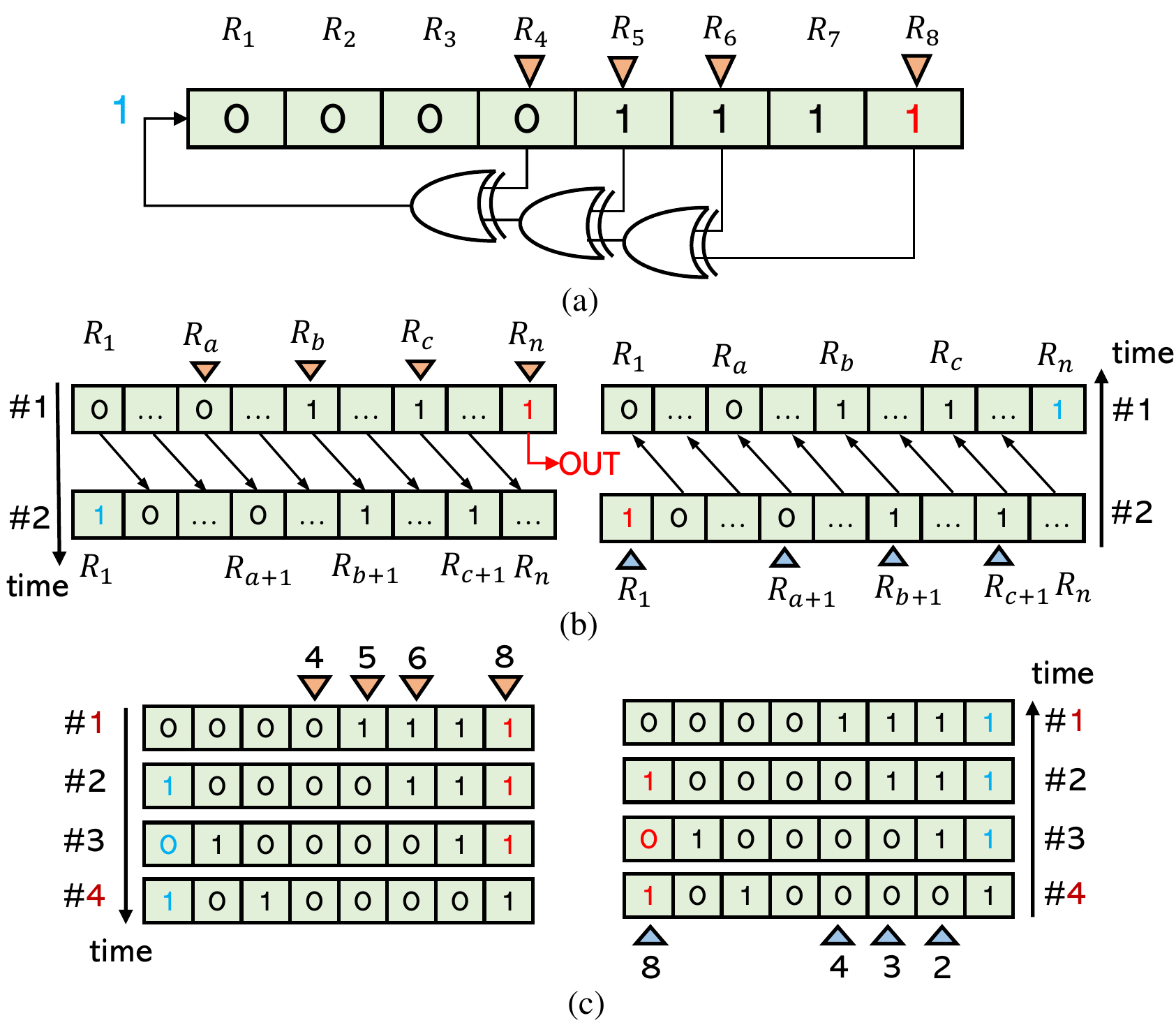}
\vspace*{-4mm}
\caption{(a) An 8-bit Fibonacci LFSR. (b) Illustration of reproducing the previous patterns by shifting the LFSR reversely. (c) Demonstration of shifting the 8-bit LFSR reversely to obtain the previous patterns in 4 cycles. Note that \#N refers to the pattern number.}
\label{fig:reverse}
\vspace*{-4.5mm}
\end{figure}

According to the Central Limit Theorem \cite{brosamler1988almost}, a binomial distribution $\mathbf{B}(n, p)$ can approximate a Gaussian distribution $\mathbf{N}(np, np(1-p))$ if $n$ is large enough. Here $n$ represents the total number of independent trials and p denotes the possibility of success for each trial. For instance, assume if there are n individual bits that have the equal possibility of being 0 or 1, the total number of ``1s" in these n bits will follow the binomial distribution $\mathbf{B}(n, 0.5)$, and further approximate the Gaussian distribution as $\mathbf{N}(0.5n, 0.25n)$ when n is large enough. Based on this insight, previous efforts \cite{kang2010fpga,cai2018vibnn,andraka1998fpga,condo2015pseudo} have proposed efficient Gaussian Random Number Generator (GRNG) by 
implementing an n-bit LFSR for uniformly distributed random bits generation and an adder tree for counting the number of ``1s". The structure of an 8-bit Fibonacci LFSR is illustrated in Fig. \ref{fig:reverse}(a). In each cycle, values in the tap registers, i.e., $R_4$, $R_5$, $R_6$ and $R_8$, are combined using three XOR gates and produce one bit to update the value in the head register $R_1$ (highlighted in blue). Meanwhile, the rest of the values shift to the neighbour register from left to right and the value in the tail register $R_8$ is dropped (highlighted in red). Through this procedure, the LFSR creates a random bit sequence named ``pattern" upon every shifting. For each pattern, the number of ``1"s are counted by the adder tree to form a Gaussian random variable (GRV). 
 
\subsection{Retrieving $\epsilon$ via Pattern Reproduction} \label{idea}
\begin{figure}[t]
\centering
\includegraphics[scale=0.66]{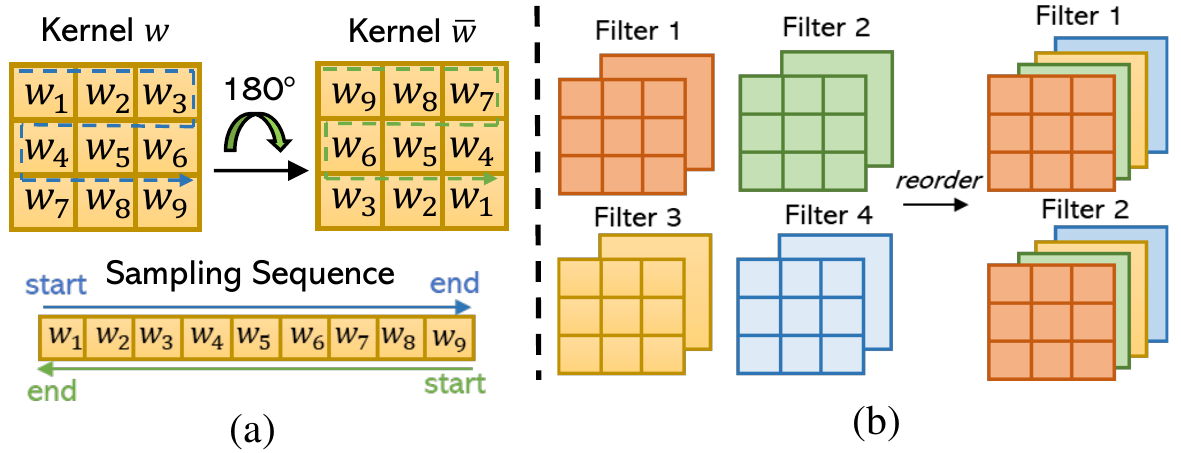}
\vspace*{-3.5mm}
\caption{(a) Kernel rotation and its relation with reversed sampling sequence. (b) Kernel reorganization.}
% Generating previous $\epsilon$s reversely (depicted as green arrow) constructs the flipped kernel without changing the computing sequence of weights (depicted as dot line)
\label{fig:rotate}
\vspace*{-4.5mm}
\end{figure}

Assume we employ one LFSR to generate $\epsilon$s for sampling all the weights during BNN training. At the Forward stage, $\epsilon$s are generated sequentially to sample from the first weight of the first layer to the last weight of the last layer, during which the LFSR continuously shifts from its initial pattern \#1 to the latest pattern \#N. At the Backward stage, we notice that the generated $\epsilon$s are requested in a reversed order, i.e., from the latest pattern \#N to the initial pattern \#1 of the LFSR, due to the two key features of the training process. At the layer-level, back-propagation executes from the last layer to the first layer, thus the $\epsilon$s generated in the last layer in the Forward stage are needed first. At the kernel-level, constructing the kernels that were rotated $180^{\circ}$ during back-propagation is equivalent to sampling the previous weights reversely (shown in Fig. \ref{fig:rotate} (a)). The aforementioned insights motivate us to reproduce the previous LFSR patterns also in a reversed order so that all the previous $\epsilon$s can be \textit{retrieved locally by LFSRs} instead of storing/fetching them during Forward/Backward stage. 

\textbf{\underline{Key design insight}.} This comes from our finding that reproducing previous LFSR patterns can be simply accomplished by shifting the current LFSR pattern in an opposite direction, combined with three XOR operations on certain registers within an LFSR, as illustrated in Fig. \ref{fig:reverse} (b). Assume a n-bit LFSR with taps $\mathbf{t}$=(a,b,c,n) is shifting right to generate the latest pattern \#2 from its initial pattern \#1. The value in the head register $R^{'}_1$ of pattern \#2 is generated by XORing the tail tap $R_n$ with other taps $R_c, R_b, R_a$ in an order:
\vspace*{-1mm}
\begin{equation}
    R^{'}_1 = ((R_n \oplus R_c) \oplus R_b) \oplus R_a 
    \label{eq4}
\end{equation}
where $\oplus$ denotes XOR operation. Meanwhile, the value in the tail register $R_n$ is dropped from the LFSR. In order to reproduce pattern \#1 from \#2, the values in $R_1,R_2...R_{n-1}$ of pattern \#1  can be obtained by left shifting pattern \#2. Now the key question is how to reproduce the value in $R_n$ of pattern \#1 since it has been dropped previously. Interestingly, for the XOR operation, one can prove that $A = C \oplus B$ if $A \oplus B = C$. Thus we rewrite Eq.\ref{eq4} in a reversed order:
\vspace*{-1mm}
\begin{equation}
    R_n = ((R^{'}_1 \oplus R_a) \oplus R_b) \oplus R_c
    \label{eq5}
\end{equation}
where $R^{'}_1$ is the head register of pattern \#2, and $R_a, R_b, R_c$ in pattern \#1 are actually $R_{a+1}, R_{b+1}, R_{c+1}$ in pattern \#2. Therefore, we can simply set $R_1, R_{a+1}, R_{b+1}, R_{c+1}$ as tap registers of pattern \#2 \textit{for the retrieval of $R_n$ in pattern \#1}, as shown in the right part of Fig. \ref{fig:reverse}(b). 
Furthermore, since the LFSR in pattern \#2 shifts reversely, the tail register $R_n$ of pattern \#2 should be updated by XORing $R^{'}_1, R_{a+1}, R_{b+1}, R_{c+1}$ of pattern \#2 orderly. In this fashion, this interesting feature can always be leveraged to retrieve the value in $R_n$ through Eq.\ref{eq5}. As can be seen, pattern \#1 is successfully retrieved from pattern \#2 via very simple logic operations. Fig. \ref{fig:reverse} (c) provides an example of reversing an 8-bit LFSR to retrieve the previous patterns.

\subsection{Potential Issues of Directly Applying LFSR Reversion to BNN Training}
\begin{figure}[t]
\centering
\includegraphics[scale=0.5]{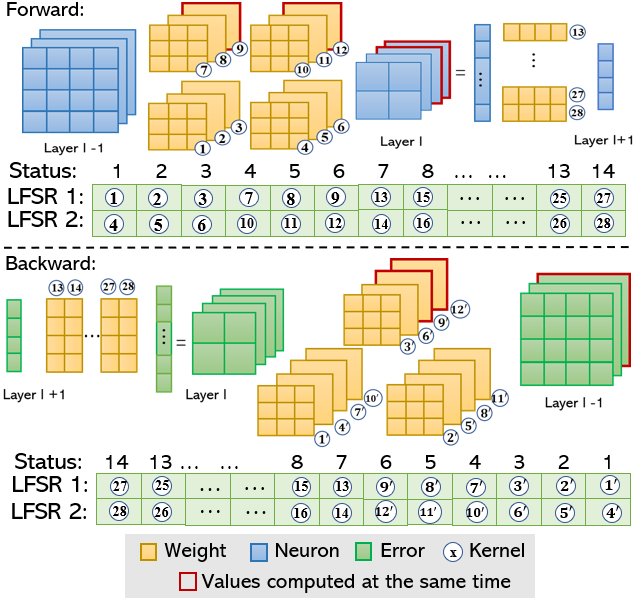}
\vspace*{-4.5mm}
\caption{An example of directly applying LFSR reversion strategy to BNN training.}
\label{fig:example}
\vspace*{-6.5mm}
\end{figure}

\begin{figure*}[t]
\centering
\includegraphics[scale=0.79]{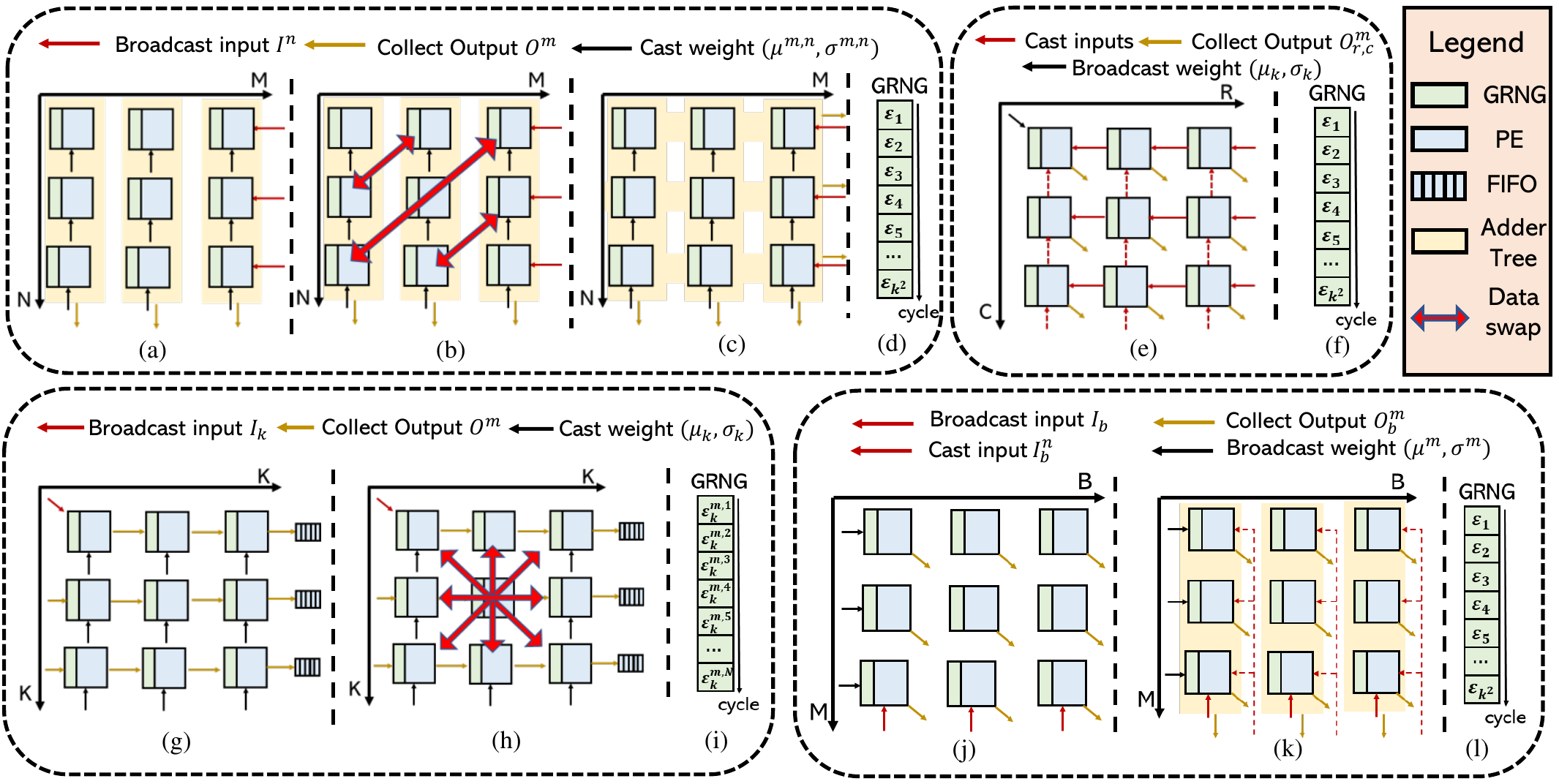}
\vspace*{-2.5mm}
\caption{(a)$\sim$(d): Basic MN-mapping, modified MN-mapping-v1, modified MN-mapping-v2 and GRNG of MN-mapping. (e)$\sim$(f): Basic RC-mapping, and GRNG of RC-mapping. (g)$\sim$(i): Basic K-mapping, modified K-mapping-v1 and GRNG of K-mapping. (j)$\sim$(l): Basic BM-mapping, modified BM-mapping-v1 and GRNG of BM-mapping.}
\label{fig:dataflow}
\vspace*{-4.5mm}
\end{figure*}

Fig. \ref{fig:example} depicts the details of applying our LFSR reversion strategy in a two-layer (convolution + fully-connected (FC)) BNN training. For simplicity of discussion, we assume two LFSRs are deployed for GRN generation. 
% For the convolutional layer, each status of a LFSR (e.g., status 1 to 6) contains 9 patterns (i.e., for a $3\times3$ kernel) and these 6 status generate $\epsilon$s for the corresponding twelve $3\times3$ kernels (from \textcircled{\small{1}} to \textcircled{\scriptsize{12}}). For the FC layer, each status of LFSR (e.g., status 7 to 14) consists of 4 patterns and these 8 status generate $\epsilon$s for the corresponding sixteen $1\times4$ weight columns (from \textcircled{\scriptsize{13}} to \textcircled{\scriptsize{28}}). 
During forward stage, for the convolutional layer, the LFSRs shift from status 1 to 6. Each status contains 9 sequential patterns to generate GRVs for a $3\times3$ kernel (each pattern per weight). For the FC layer, the LFSRs continue shifting from status 7 to 14. Each status contains 4 sequential patterns for a $1\times4$ weight vector.
% The two LFSRs shift from the initial status 1 to the last status 14 after the Forward stage. 
During the Backward stage, by shifting the LFSRs reversely, all the previous status are retrieved in a reversed sequence that satisfies the weight fetching request by backpropagation. Note that for convolutional layers, the flipped ($180^{\circ}$ rotated) kernels \textcircled{\scriptsize{$x'$}} can be constructed by the reversed order of \textcircled{\scriptsize{$x$}} according to Fig. \ref{fig:rotate}(a). And for the FC layers, since the internal weight order of each weight column (e.g.,$1\times4$ matrix) is not altered, the original weight matrices can all be retrieved via LFSR reversion. However, as shown in Fig.\ref{fig:rotate} (b), since the kernels are reorganized across the input channel (N) dimension and output channel (M) dimension during the Backward stage, the computation flow could become inconsistent with that in the Forward stage. For example, at status 6 during the Forward stage in Fig.\ref{fig:example}, the partial sums calculated by kernel \textcircled{\scriptsize{$9$}} and \textcircled{\scriptsize{$12$}} are \textit{accumulated separately} for the last two output channels (i.e., the blue blocks highlighted by red at layer $l$). When applying our LFSR reversion, kernel \textcircled{\scriptsize{$9^{'}$}} and \textcircled{\scriptsize{$12^{'}$}} will be constructed at status 6 during Backward stage. At this time, instead of being accumulated separately, the partial sums calculated by kernel \textcircled{\scriptsize{$9^{'}$}} and \textcircled{\scriptsize{$12^{'}$}} are \textit{added together} for one single output channel (i.e., the green block highlighted by red at layer $l-1$). Although our LFSR reversed shifting can still retrieve all the $\epsilon$s, such computation inconsistency between the Forward and Backward stages may pose significant design inefficiency for training accelerator design. Furthermore, this factor complicates the design choice selection due to the unclear impact our LFSR reversion strategy may pose on accelerators that adopt different computation mapping schemes. Thus, it is important to first understand the accelerator design space for our shift-BNN.   

\section{Design Space Exploration} \label{discuss}

As discussed in Section \ref{cf} (also see Fig. \ref{fig:bg} (b)), processing a typical DNN layer during any training stage can be decomposed into a six-dimension for-loop execution. Instead of executing each dimension sequentially, the state-of-the-art DNN accelerators usually select several dimensions and compute them simultaneously, during which MACs along a certain dimension are mapped onto a group of Processing Elements (PEs) that operate in parallel. Choosing different mapping dimensions creates a significant divergence in design efficiency. Generally, there have been three major types of computation mapping strategies for DNN \textit{inference}: kernel (K-dimension) mapping, e.g., systolic array \cite{farabet2009cnp}, input channel and output channel (MN-dimension) mapping, e.g., Diannao \cite{chen2014diannao}, NVDLA \cite{nvdla}, and output feature mapping (RC-dimension) mapping, e.g., Shidiannao \cite{du2015shidiannao}. Since DNN \textit{training} could also perform mini-batch processing, a batch and output channel (BM-dimension) mapping method \cite{yang2020procrustes} is also under consideration. To efficiently apply our design insights into BNN training, we comprehensively study the impact of our LFSR reversed shifting strategy on the four types of state-of-the-art computation mappings to explore the design space for BNN training accelerator. Specifically, we qualitatively discuss the design possibility by using each mapping,and finally select the optimal mapping to support our proposed Shift-BNN design. 
In the following analysis, we apply superscript $m$ and $n$ to denote the index of output and input channel, and subscript $k$, $(r,c)$ and $b$ to denote the weight location inside a kernel, the neuron/error location on an output feature map and the index of a training example in a mini-batch, respectively. 
% The missing superscripts/subscripts for a certain variable (e.g., weight, neuron, etc.) means the corresponding dimensions are insignificant for the discussion.

% including weight parameter $(\mu, \sigma)$, Gaussian random number $\epsilon$, input neuron $I$ and output neuron $O$ 

\textbf{\underline{MN-dimension mapping}.}
Fig. \ref{fig:dataflow} (a) illustrates a basic architecture for MN-dimension mapping. The x-axis of the 2-D PE array (we assume the size is $3\times3$ for simplicity) represents M-dimension mapping and the y-axis represents N-dimension mapping. As BNN training demands weight sampling, a GRNG is attached to each PE to generate $\epsilon$s for weight parameters $(\mu, \sigma)$, which will be the common case among all four types of mapping methods. In each cycle, an input neuron $I^n$ from a certain input channel $n$ broadcasts horizontally to a row of PEs, where each PE calculates the partial sums for a certain output channel $m$. These partial sums are collected vertically by an adder tree (denoted by the yellow bar) and summed up until a output neuron is generated. In this scheme, a PE located at coordinate $(m,n)$ will require a $K \times K$ kernel from input channel $n$ and output channel $m$ to produce the partial sum of an output neuron. Therefore, during FW, the LFSR in each GRNG generates $\{\epsilon_1, \epsilon_2,...,\epsilon_{K^2}\}$ sequentially to produce a sampled kernel $\{w_1, w_2,...,w_{K^2}\}$, as shown in Fig. \ref{fig:dataflow} (d). With the proposed LFSR reversion strategy, the flipped kernel can be reconstructed by shifting the LFSR reversely during BW. However, also during this stage, as the kernels are also reorganized in the MN-dimension, the partial sums generated in PE rows should be summed up instead of being accumulated separately (Sec.4.3). This results in the inconsistent computation patterns between FW and BW. To address this inconsistency in a uniform architecture design, one possible solution is to swap the Gaussian random variables, i.e., $\epsilon$s, between PE $(m,n)$ and PE $(n,m)$ and then load the corresponding weight parameters and input neurons during the BW stage, as shown in Fig. \ref{fig:dataflow} (b). Nevertheless, such design requires extra interconnections between PEs, leading to $\mathcal{O}(n^2)$ wiring overhead for a $n \times n$ PE array, which hinders design scalability. Moreover, there must be an equal number of PEs in a row and a column due to the swapping mechanism, which further limits the design flexibility. Fig. \ref{fig:dataflow} (c) shows an alternative design that avoids the data communication between PEs. In this design, during BW the partial sums generated by a PE row are summed up to an output neuron with duplicated adder trees. The partial sums generated by a PE column are accumulated separately by directly sending each of them to the output buffer. However, this method still requires an $n$-input adder tree for each row of PEs, which incurs extra resource and energy overheads. 

\textbf{\underline{RC-dimension mapping}.}
% \begin{figure}[t]
% \centering
% \includegraphics[scale=0.6]{figs/rc-dataflow-v2.PNG}
% \vspace*{-2.5mm}
% \caption{(a) Basic output feature map (RC) dimension mapping strategy. (b) The sequence of generated $\epsilon$s in each GRNG.}
% \label{fig:rc}
% \vspace*{-4.5mm}
% \end{figure}
Fig. \ref{fig:dataflow} (e) shows the basic output feature map (RC) dimension mapping strategy, where neurons on a $R \times C$ output feature map are mapped to a $R \times C$ 2-D PE array and computed simultaneously. In each cycle, one weight from a $K \times K$ kernel is broadcast to all PEs while a group of new input neurons are fed to the rightmost (or bottom) PEs. The partial sums stay in the PE and are accumulated to generate the output neurons as the input neurons flow from right to left (or bottom to up) through the PE array. Since the weight is fetched sequentially from a $K \times K$ kernel, the GRNG also produces $\{\epsilon_1, \epsilon_2,...,\epsilon_{K^2}\}$ during FW. Thus, the flipped kernels can be reconstructed by shifting LFSR reversely during BW. Furthermore, since RC-dimension mapping is irrelevant with M- or N-dimension parallelism, it will not suffer from the $\epsilon$ swapping issue from MN-mapping. Nevertheless, kernel reorganization still has a slight impact on RC-mapping. During the FW stage, since the kernels are fetched along the N-dimension first and then M-dimension, the partial sum of an output neuron is accumulated inside the PE continuously until the output neuron is generated. However, during the BW stage, the kernels are fetched along the M-dimension first and then N-dimension; so the partial sum of an output neuron is sent to the output buffer and waits to be read and accumulated in the PE intermittently. Therefore, two types of control modes are required in RC-mapping. 
%\begin{figure}[t]
% \centering
% \includegraphics[scale=0.62]{figs/k-dataflow.PNG}
% \vspace*{-2.5mm}
% \caption{(a) Basic Kernel (K) dimension mapping strategy. (b) K-dimension mapping v1. (c) The sequence of generated $\epsilon$s in each GRNG.}
% \label{fig:kk}
% \vspace*{-4.5mm}
% \end{figure}

\begin{figure*}[t]
\centering
\includegraphics[scale=0.56]{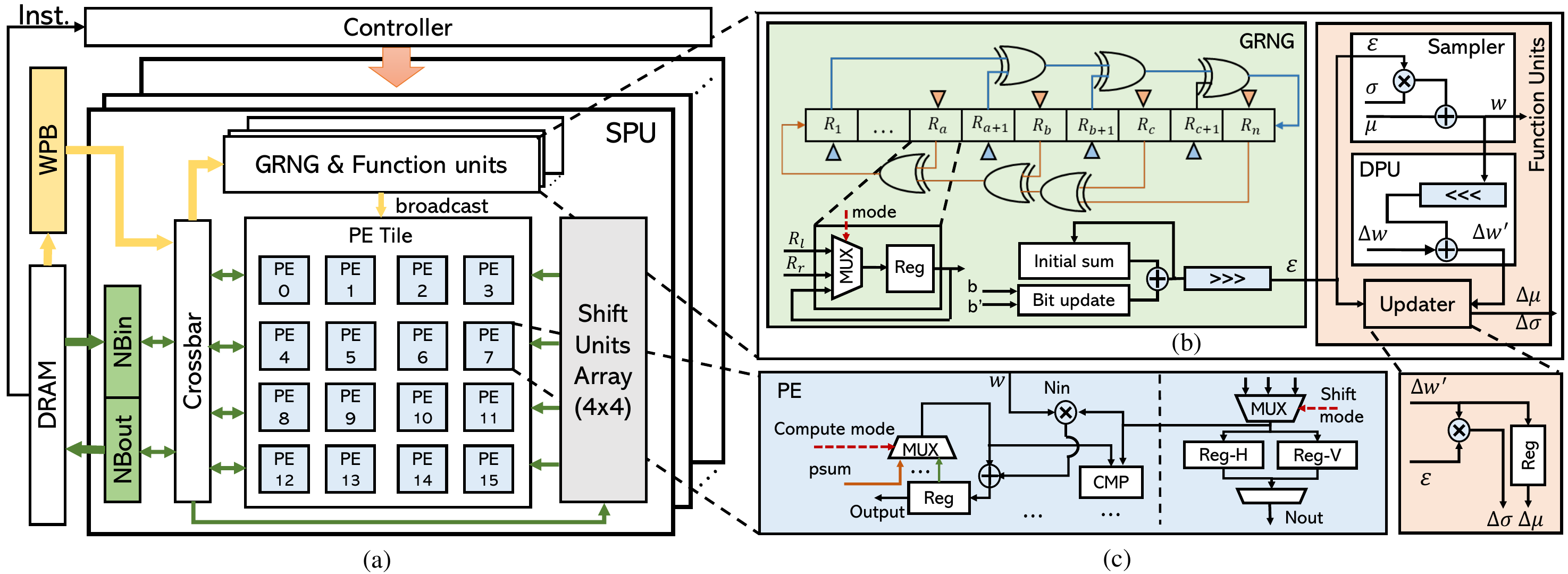}
\vspace*{-3.5mm}
\caption{(a) Overview of our proposed Shift-BNN training accelerator. (b) The microarchitecture of GRNG and function units. (c) PE implementation for RC-mapping computation flow.}
\label{fig:archit}
\vspace*{-4.5mm}
\end{figure*}

\textbf{\underline{K-dimension mapping}.}
Fig. \ref{fig:dataflow} (g) shows the basic kernel (K) dimension mapping method, where a $K \times K$ kernel is mapped to a $K \times K$ 2-D PE array and stays until all the computation related to that kernel is completed. In each cycle, an input neuron is broadcast to all the PEs and multiplied with $K \times K$ weights inside a kernel. The partial sums are propagated and accumulated through the PEs to generate the output neurons. Under this scheme, during FW the PE array requires the kernel from the next input channel when the computation of the current kernel is finished. Hence, the GRNG generates $\epsilon$s for weights along the N-dimension sequentially from the first to the last input channel, i.e., $\{\epsilon_k^{m,1}, \epsilon_k^{m,2},...,\epsilon_k^{m,N}\}$, as shown in Fig. \ref{fig:dataflow} (i). During BW, reverse shifting LFSR can retrieve the original kernels from the last to the first input channel. However, K-dimension mapping can not reorder the weights to construct the flipped kernels required by the BW stage as the weights inside a kernel are sampled simultaneously. In fact, due to the kernel flipping, the $\epsilon$ generated by a certain PE during FW is required by another PE during BW. Fig. \ref{fig:dataflow} (h) illustrates a solution for K-dimension mapping: adding datapaths between PEs for $\epsilon$ swapping. However, similar to the MN-dimension-v1 (as shown in Fig.\ref{fig:dataflow} (b)), this design causes $\mathcal{O}(n^2)$ wiring overhead for a $n \times n$ PE array. Moreover, due to the kernel reorganization, K-mapping also requires two types of control modes for different accumulation manners.

\textbf{\underline{BM-dimension mapping}.}
% \begin{figure}[t]
% \centering
% \includegraphics[scale=0.62]{figs/bm-dataflow.PNG}
% \vspace*{-2.5mm}
% \caption{(a) Basic batch and output channel (BM) dimension mapping strategy. (b) BM-dimension mapping v1. (c) The sequence of generated $\epsilon$s in each GRNG.}
% \label{fig:bm}
% \vspace*{-4.5mm}
% \end{figure}
Fig. \ref{fig:dataflow} (j) illustrates the basic batch and output channel (BM) dimension mapping strategy, where the horizontally distributed PEs are processing different training examples and the vertically distributed PEs are calculating neurons in different output channels separately. In each cycle, a pair of weight parameters $(\mu^m, \sigma^m)$ from a certain output channel $m$ is broadcast to an entire row of PEs while an input neuron $I_b$ from a certain training example $b$ is broadcast to an entire column. The output neurons can be collected in each PE. As the weights inside a certain kernel are requested sequentially (shown in Fig. \ref{fig:dataflow} (l)), LFSR reversion can help reconstruct the flipped kernels. However, due to the kernel reorganization, the reconstructed kernels in a column of PEs should be used for N-dimension computation instead of M-dimension computation. Specifically, at the BW stage, the partial sums generated by PE columns should be summed up instead of being accumulated separately. To address this issue, an additional n-input adder tree is required for each PE column. Meanwhile, different input neurons $I_b^n$ from input channel $n$ are sent to each PE column, resulting in two different input buffer designs (Fig. \ref{fig:dataflow} (k)). Therefore, this architecture not only incurs large hardware overhead but also leads to high design complexity.

In conclusion, the RC-dimension mapping strategy (Fig. \ref{fig:dataflow} (e)) only incurs modest design overhead compared to the other three mapping methods when applying our LFSR reversion strategy, which makes it an ideal fundamental computation mapping for designing our Shift-BNN architecture.

\section{Shift-BNN Architecture Design} \label{archi}
%\subsection{Hardware architecture of Shift-BNN}
%We propose a highly customized hardware design for Shift-BNN that integrates our LFSR reversion strategy with the basic architecture of RC-dimension mapping.

\subsection{Architecture Overview}
Figure \ref{fig:archit} illustrates the overall architecture of our proposed Shift-BNN training accelerator, which comprises of a 3D PE array distributed to 16 Sample Processing Units (SPUs), a weight parameter buffer (WPB), and a central controller. Each SPU consists of an input/output neuron buffer (NBin/NBout), 16 slices of GRNG and function units, a $4 \times 4$ PE tile, a 4 $ \times 4$ array of shift units, and a crossbar. Following the aforementioned LFSR reversion technique and the computation mapping consideration, our accelerator presents the following features: (1) \textit{a hybrid dataflow} that adopts RC-dimension on 2D PE tiles and sample-level parallelism across SPUs, both of which exploit significant opportunities for data reuse; (2) \textit{an efficient GRNG design} which can generate Gaussian random variables $\epsilon$s sequentially during FW stage and reproduce the previous $\epsilon$s reversely during BW stage; (3) \textit{function units} design that satisfies necessary mathematical operations, i.e., weight sampling, derivative calculation of prior and posterior, and weight updating during the BNN training; (4) \textit{light implementation} of RC-dimension mapping architecture by using a PE tile, an array of shift units and a crossbar.
\vspace*{-2mm}
% To improve the resource utilization and reduce the weight transfer of original 2D-mapping dataflow, we add one dimension to the architecture to support parallel computations for different samples.

\subsection{SPUs and Dataflow}
Since the weight parameters $(\mu, \sigma)$ are shared among sampled models, it is natural to process a batch of sampled models in parallel to increase the data reuse of weight parameters. Our design leverages such opportunities by allocating the workloads of training each sampled model to an individual SPU, which operates independently and in parallel with other SPUs. Each SPU is further equipped with the RC-dimension mapping scheme that maximizes the data reuse of input neurons on a 2D feature map. We describe the main features of an SPU as follows.

\textbf{\underline{PE tile, shift unit and crossbar}.} All convolution operations are performed in the 2D PE tile during all three stages of BNN training (i.e., FW, BW and GC). For simplicity of discussion, we use the FW stage as an example to illustrate the datapath design and the computation flow.
% For simplicity, we uniformly use ``synapse" and ``input" to represent weight (in FW and BW)/error (in GC) and input feature map (in FW and GC)/error (in BW), respectively. 
Fig. \ref{fig:archit} (a) shows the datapath for a convolutional layer, in which a sampled weight from the GRNG \& function units is broadcast to all the PEs and multiplies with the input neuron, which will shift to the left (or up) neighbour PE in the next cycle (Fig.\ref{fig:dataflow}(e)-(f)). To support this type of dataflow, a dedicated PE design is implemented upon a typical inference accelerator \cite{du2015shidiannao} that adopts RC-dimension mapping, shown in Fig.\ref{fig:archit} (c). The right part of the PE is a shift unit. It determines which input neuron (Nin) should be received by the PE and which neuron that is stored in Reg-H/Reg-V should be sent (Nout) to the other PEs. The selected input neuron and the broadcast weight will then enter into the computation unit, which is depicted at the left part of the PE and performs basic MAC operations, ReLU functions and max pooling operations to produce the output neurons. Importantly, due to the kernel reorganization and $\epsilon$ reproducing technique at the BW stage (Sec.\ref{discuss}), our PE design supports \textit{two types of accumulation modes}. (1) During the FW stage, since the kernels are fetched along the N-dimension first and then M-dimension, the partial sum is repeatedly fetched back to the PE, depicted by the green arrow in Fig. \ref{fig:archit} (c). (2) During BW stage, the kernels are fetched along the M-dimension first and then N-dimension, thus the partial sum (named psum in the figure) is fetched from NBout and then gets accumulated in the PE intermittently, depicted by the orange arrow in Fig. \ref{fig:archit} (c). Our PE design switches between these two accumulation modes for FW and BW stages. 
Furthermore, to satisfy the complex data requests from the PE tile, a crossbar is inserted between WPB, NBin, NBout and PE tile to select the appropriate data read from the buffer. Additionally, instead of using a column buffer in \cite{du2015shidiannao}, we employ a light-weight $4 \times 4$ shift units array which stores the candidate input neurons that the PE tile will need in the next four cycles. The array is organized in the same way as the PE tile spatially and each shift unit is actually the same as the right part of the PE for simple data shifting operations.% which is shown in Fig. \ref{fig:archit} (d). 

\textbf{\underline{Efficient GRNG design}.} A SPU contains $4 \times 4$ GRNGs, which corresponds to the $4 \times 4$ PE tile.
For a convolutional layer, since one weight is shared by every PE, only one GRNG needs to be enabled to generate one $\epsilon$ at a time. While for a FC layer, PEs require different sampled weights from the GRNG \& function units thus all GRNGs are enabled to provide $\epsilon$s to sample weights for their corresponding PE.  Fig. \ref{fig:archit} (b) left illustrates the microarchitecture of a single GRNG which consists of a 256-bit LFSR and an $\epsilon$ generator. The GRNG features two properties. Firstly, it possesses three operating modes. (1) The forward mode for FW stage, during which the LFSR shifts from left to right. Each register (except $R_1$) of LFSR receives the values from the left neighbour register (named $R_l$) while $R_1$ gets updated by the orange taps. (2) The backward mode for BW stage, during which the GRNG switches to the reverse mode and shifts from right to left. Each register (except $R_n$) of LFSR receives the values from the right neighbour register (named $R_r$) while $R_n$ gets updated by the blue taps. %As illustrated in Section \ref{idea}, the LFSR will reproduce all the previous patterns in the reverse mode. 
(3) The idle mode, during which registers in the LFSR receive their own values and will not be updated. Secondly, since counting the number of ``1s" (or the sum) of a LFSR pattern with an adder tree may cause large overhead \cite{cai2018vibnn}, the proposed $\epsilon$ generator uses a more efficient way to generate $\epsilon$s based on the LFSR patterns. Specifically, we store the sum of the bits in the LFSR's initial seed in a register and track the difference between the old value ($b$) and the updated value ($b^{'}$) at $R_1$ or $R_n$ depending on the operating mode. The difference, i.e., bit update, will be added to the initial sum to form the current sum of LFSR which are then used to update the register of the initial sum. 
% \textcolor{red}{In order to obtain the unit Gaussian random variables $\epsilon$s that strictly follow $\mathbf{N}(0,I)$, the current sum of the LFSR that follows $\mathbf{N}(128, 64)$ should be shifted and scaled properly. We first shift the mean value of 128 to 0 by deducting 128 from the initial sum at the beginning and then scale the standard deviation from 8 to 1 by right shifting the current sum by 3 bits.}

\textbf{\underline{Function units}.}
The function units consist of a sampler, a derivative processing unit (DPU), and a weight parameter updater. As a whole, the function units receive the $(\mu, \sigma)$ and $\epsilon$ from the crossbar and the GRNG respectively, and accomplish two tasks: weight sampling and final gradient calculation of the weight parameters. During both FW and BW stages, the weight sampling is performed in a sampler that applies the weight parameters $(\mu, \sigma)$ to the Gaussian random number $\epsilon$ using a multiplier and an adder. The produced weight is sent to the PE tile and the DPU. During the BW stage, the DPU and the updater are both activated. The DPU uses the received reconstructed weight to calculate the derivatives of the sum of the prior and posterior with respect to the weight, $\Delta w_p$. By decomposing the prior and posterior terms into a log form, $\Delta w_p$ can be approximated as $\frac{w}{\sigma^2_c}$. Since $\sigma_c$ is a constant value of prior distribution and is usually chosen as 0.5, we thus calculate the $\Delta w_p$ by left shifting $w$ 2 bits. The $\Delta w_p$ is then added to the gradient of likelihood computed in the GC stage to obtain the final gradient $\Delta w^{'}$. Lastly, in order to update the weight parameters, the updater calculates the gradients of $(\mu, \sigma)$ using $\Delta w^{'}$ and $\epsilon$, which corresponds to the process \textcircled{\small{3}} in Fig.\ref{fig:bg} (a). The produced $(\Delta\mu,\Delta\sigma)$ will be further averaged across different SPUs and then used to update the weight parameters. 

\textbf{\underline{Buffer design}.}
To support the dataflow of RC-mapping in an SPU, we follow a similar design principle in \cite{du2015shidiannao} to organize the data in the neuron buffer,i.e., NBin/NBout. NBin/NBout comprises multiple banks. Each bank provides the neurons requested by a PE row through the crossbar. For the weight parameter buffer (or \textit{WPB}), we split it into two sub-buffers that store $\mu$ and $\sigma$ separately. Each sub-buffer is also designed to consist of multiple banks and each entry of the bank stores the weights for a PE row. For a convolutional layer, one weight parameter is selected by the crossbar at each cycle while for an FC layer the entire entry read from the bank is sent to a PE row. Note that although the convolution operands (e.g., weight, neuron, error, and gradient) vary across the three BNN training stages, our uniform design of data organization in WPB, NBin and NBout is beneficial for the buffer function swapping. For example, during the BW stage, the error feature maps of layer $l+1$ stored in NBout can serve as the weights for the gradient calculation of layer $l$ by temporarily treating the NBout as WPB.

% \textbf{\underline{Handling various sample sizes}.}
% Shift-BNN employs 16 SPUs to concurrently process 16 sampled models. In some cases, when the demand for model robustness is relaxed, the number of sampled models can be smaller than 16. This may lead to SPU idleness. To tackle this, we simply allocate the sampled models of the next training example to the idle SPUs since: (1) the training examples in a mini-batch can be processed simultaneously; and (2) different training examples also share the same weight parameters $(\mu, \sigma)$. Additionally, the gradients of weight parameters, i.e., $(\Delta\mu,\Delta\sigma)$, can be still aggregated across the SPUs to calculate the average loss for the mini-batch without any architecture modification. Additionally, our design also shows good scalability for large sample sizes, which will be discussed in Sec.\ref{scalability}.

\section{Evaluation} \label{exp}
\subsection{Experimental Methodology} \label{setup}
\textbf{BNN models and training datasets.}
We evaluate Shift-BNN by training on five representative BNN models. Among them, B-MLP \cite{cai2018vibnn} (fully-connected BNN with 3 hidden layer) is trained with MNIST \cite{deng2012mnist}. B-LeNet (built on LeNet\cite{lecun2015lenet}) is trained with CIFAR-10 \cite{krizhevsky2009learning}. These two networks are mostly adopted to handle small but safety-critical tasks. B-AlexNet (built on AlexNet\cite{krizhevsky2012imagenet}), B-VGG (built on VGG16 \cite{simonyan2014very}) and B-ResNet (built on ResNet-18 \cite{he2016deep}) are trained with ImageNet datasets \cite{deng2009imagenet}, which are used to deal with more complex tasks in the unfamiliar environments. For generality, the BNN models are trained with various number of samples, e.g., 8, 16, 32, 64, and 128 (if needed) samples.

\textbf{Comparison cases.}
To demonstrate the effectiveness of Shift-BNN, we compare it with three training accelerators: Firstly, since Shift-BNN adopts RC-mapping as the fundamental design strategy, we compare it with the \textit{RC-accelerator} that adopts RC-mapping strategy but without LFSR reversion technique. Secondly, since MN-mapping is commonly used in existing DNN training accelerators \cite{mahmoud2020tensordash,zhang2019eager}, we employ an \textit{MN-accelerator} that adopts MN-mapping strategy without LFSR reversion technique as the baseline accelerator for generality, which is also used for our preliminary investigation in Sec.\ref{challenge}. Thirdly, to verify the analysis about design alternatives (see Sec.\ref{discuss}), we further test the effectiveness of our LFSR reversion strategy on MN-accelerator by comparing with an \textit{MN-Shift-accelerator} that adopts both MN-mapping strategy and LFSR reversion technique. To overcome the challenges caused by our LFSR reversion to the MN-mapping scheme, we follow the design principle in Fig. \ref{fig:dataflow} (c). For fair comparison, all accelerators employ 16 $4 \times 4$ PE tile and are allocated with on-chip buffer of the same size. The 16 PE tiles process 16 sampled models simultaneously for the same extent of weight parameter reuse. We evaluate the energy efficiency (performance/power) of Shift-BNN and compare with the modern GPU, i.e., Nvidia Telsa P100. We use Pytorch \cite{paszke2019pytorch} to implement and train the BNNs from scratch, and the training hyperparameters (e.g., batch size, epochs. etc) are kept the same as in other comparison cases. The execution latency and energy consumption are extracted from the GPU runtime information obtained by Nvidia Profiler \cite{nvprof}.

\textbf{Experimental Setup.}
All accelerator designs are implemented in Verilog RTL and synthesized on a Xilinx Virtex-7 VC709 FPGA evaluation board. For off-chip memory access, the accelerators communicate with two sets of DDR3 DRAM that provide sufficient data transfer rate to the PE tiles via a Memory Interface Generator \cite{ddr3}. The execution time results are obtained from the post-synthesis design and the energy consumption is further evaluated with Xilinx Power Estimator (XPE) \cite{xpe}. The data precision for all architectures is set to 16-bit and the operating frequency is set as 200MHz. 

\begin{figure}[t]
\centering
\includegraphics[height=36mm,width=75mm]{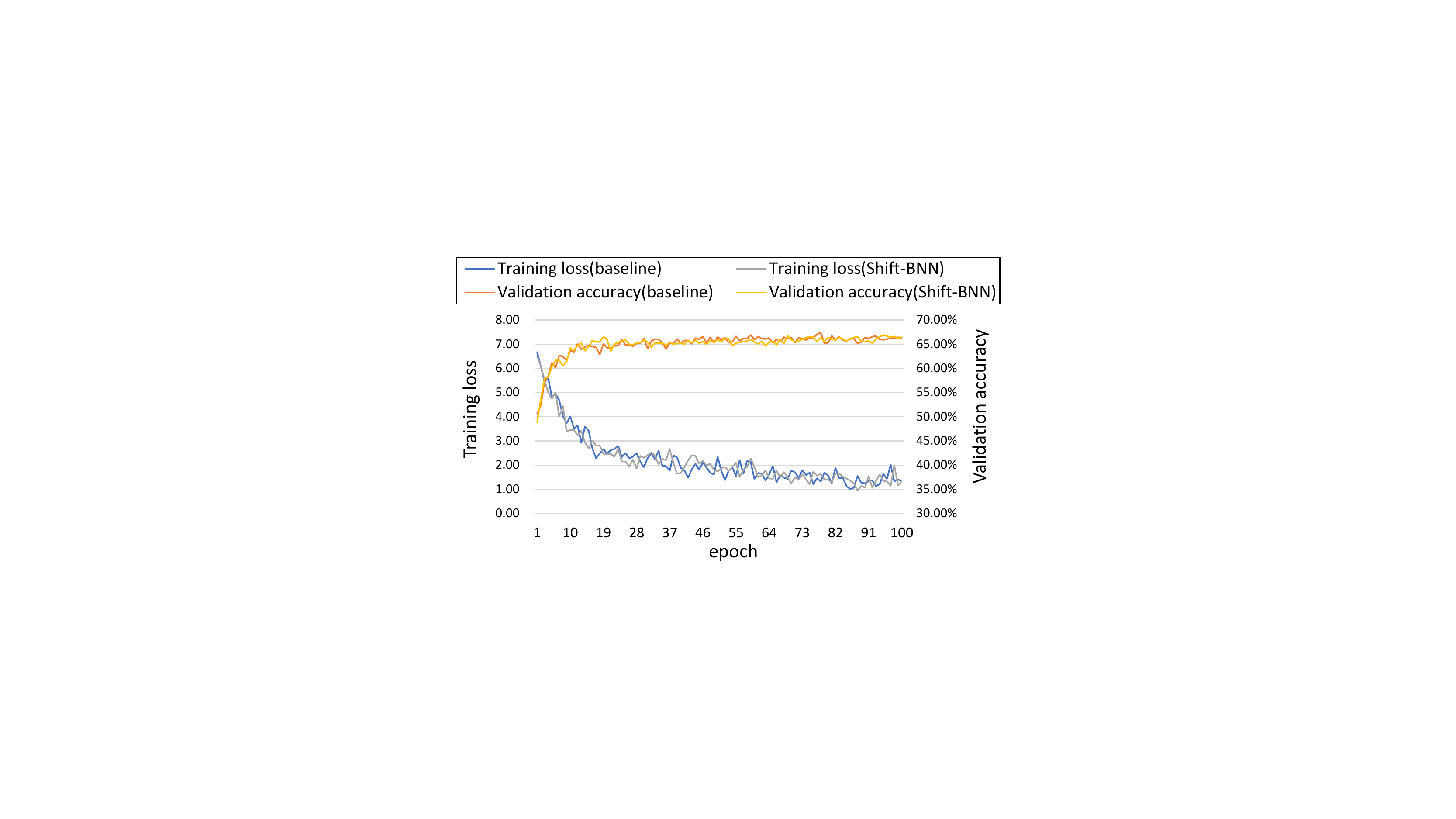}
\vspace*{-4.5mm}
\caption{Validation accuracy and training loss over training time for Shift-BNN and vanilla BNN training algorithm on B-LeNet trained with CIFAR-10.}
\label{fig:curve}
\vspace*{-1.5mm}
\end{figure}

\textbf{Training quality.}
Figure \ref{fig:curve} compares the training curve of B-LeNet when using the vanilla BNN training algorithm on Pytorch (baseline) and Shift-BNN. The training hyperparameters and data type are kept the same in baseline and Shift-BNN. It can be seen that Shift-BNN does not affect the overall training iterations to convergence and the final accuracy. Similar behavior is observed on the other networks. This is because our LFSR reversion strategy fundamentally does not modify the training algorithm and simply manages to accurately retrieve all the $\epsilon$s during the entire training process. Hence, we only evaluate and validate the training quality results on Shift-BNN when using different bit length. Table \ref{table:acc} shows how different bit lengths affect the validation accuracy of five BNN models. The accuracy results are obtained after the same training epochs for a certain network. As can be seen, training with 16-bit precision only brings an average 0.31\% accuracy drop compared with single-precision training. This negligible loss may be due to the error tolerance nature of the sampling process during BNN training. While Shift-BNN can employ 32-bit floating point arithmetic to achieve lossless training, the lower precision training is more attractive as lower precision computation potentially consumes much less energy.  
\vspace*{-1mm}
\begin{table}[h]
\vspace*{-0.5mm}
\caption{Data type vs validation accuracy.}
  \vspace*{-1.5mm}
  \scriptsize
  \centering
  \begin{tabular}{|c|c|c|c|c|c|}
    \hline
    \textbf{Network} & \textbf{B-MLP} & \textbf{B-LeNet} & \textbf{B-AlexNet} & \textbf{B-VGG} & \textbf{B-ResNet} \\
    \hline
    Dataset & MNIST & CIFAR-10 & ImageNet & ImageNet & ImageNet \\
    \hline
    Val-acc(8b) & 95.67\% &	62.80\%	& NaN$^1$ & 45.50\% & NaN \\
    \hline
    Val-acc(16b) & 98.05\% & 65.62\% & 59.95\% & 67.52\% & 68.12\%	\\
    \hline
    Val-acc(32b) & 98.11\% & 65.81\% & 60.10\% & 67.76\% & 69.03\%  \\
    \hline
  \end{tabular}
  \vspace*{-2.5mm}
  \label{table:acc}
\end{table}

\begin{scriptsize}
{$^1$The network hardly converges due to the low precision 8-bit BNN training.}
\end{scriptsize}

\subsection{Evaluation Results}

\begin{figure}[t]
\centering
\includegraphics[scale=0.58]{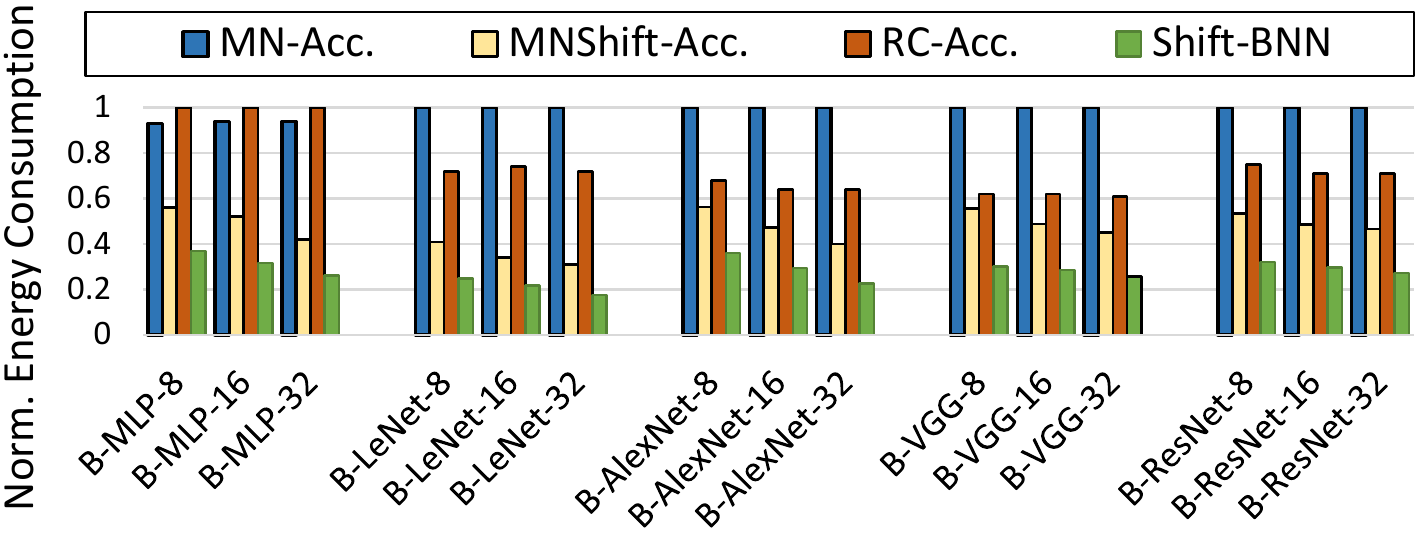}
\vspace*{-4.5mm}
\caption{Energy consumption comparison between Shift-BNN and other designs.}
\label{fig:energy}
\vspace*{-2.5mm}
\end{figure}

\begin{figure}[t]
\centering
\includegraphics[scale=0.57]{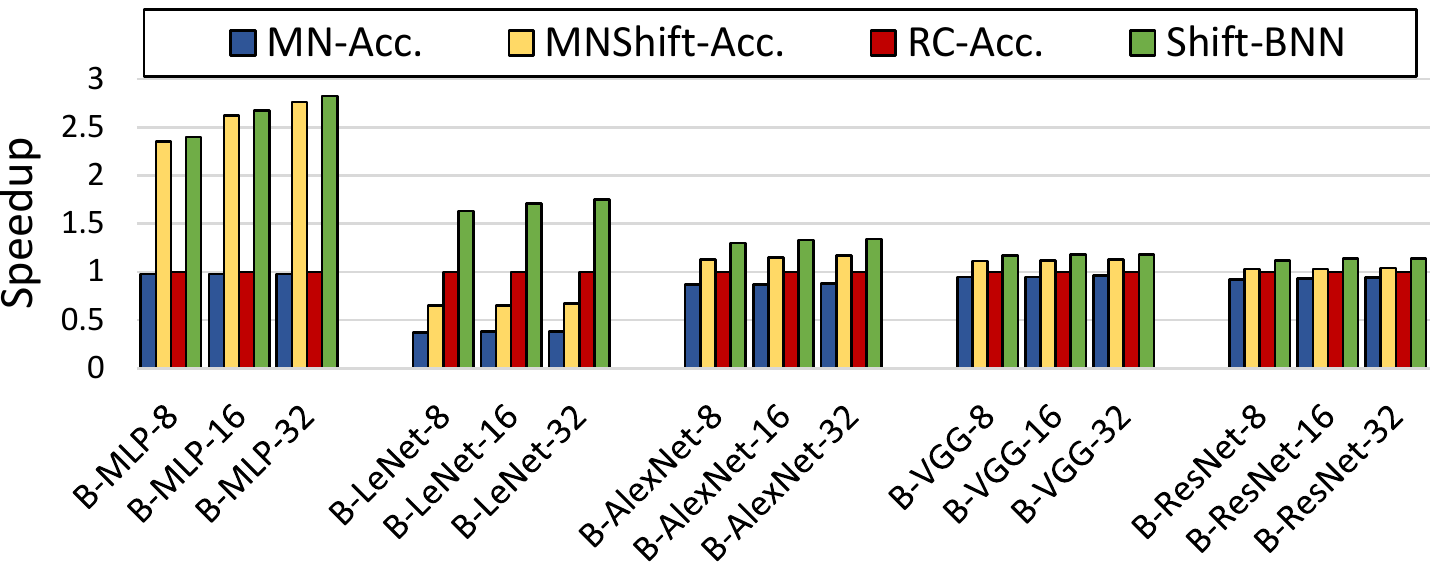}
\vspace*{-3.5mm}
\caption{The speedup of Shift-BNN compared with other accelerators.}
\label{fig:speedup}
\vspace*{-3.5mm}
\end{figure}

\begin{figure}[t]
\centering
\includegraphics[scale=0.59]{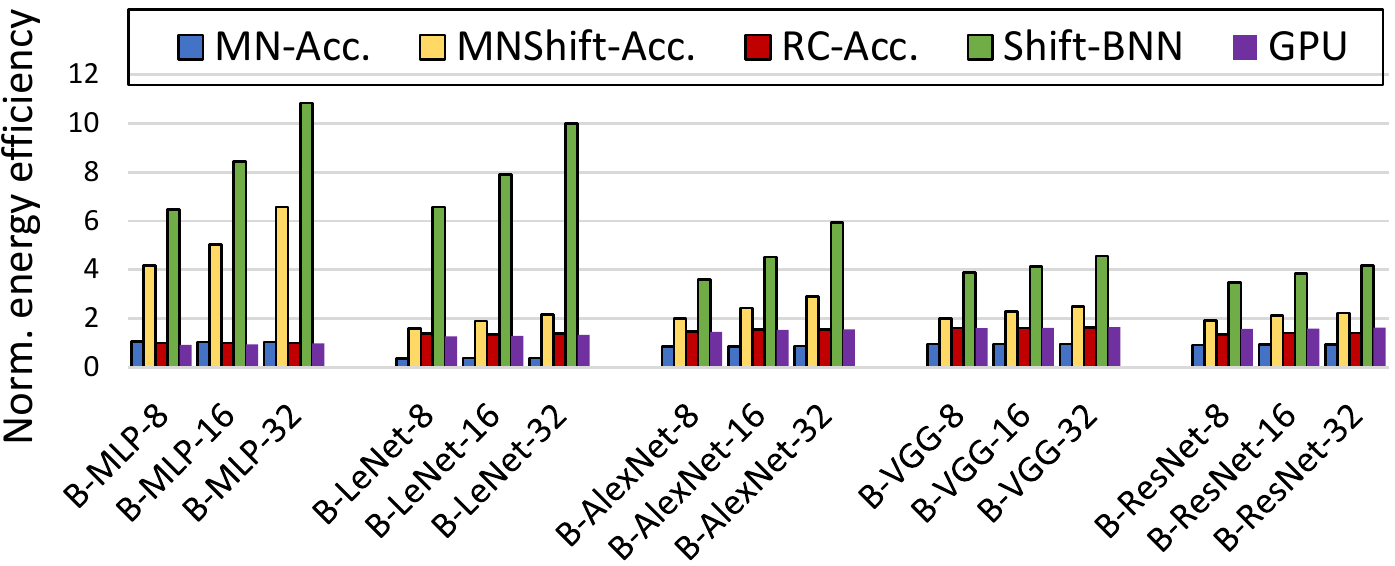}
\vspace*{-4.5mm}
\caption{Energy efficiency comparison between Shift-BNN and other designs.}
\label{fig:ee}
\vspace*{-5.5mm}
\end{figure}

\begin{figure*}[t]
\centering
\includegraphics[scale=0.59]{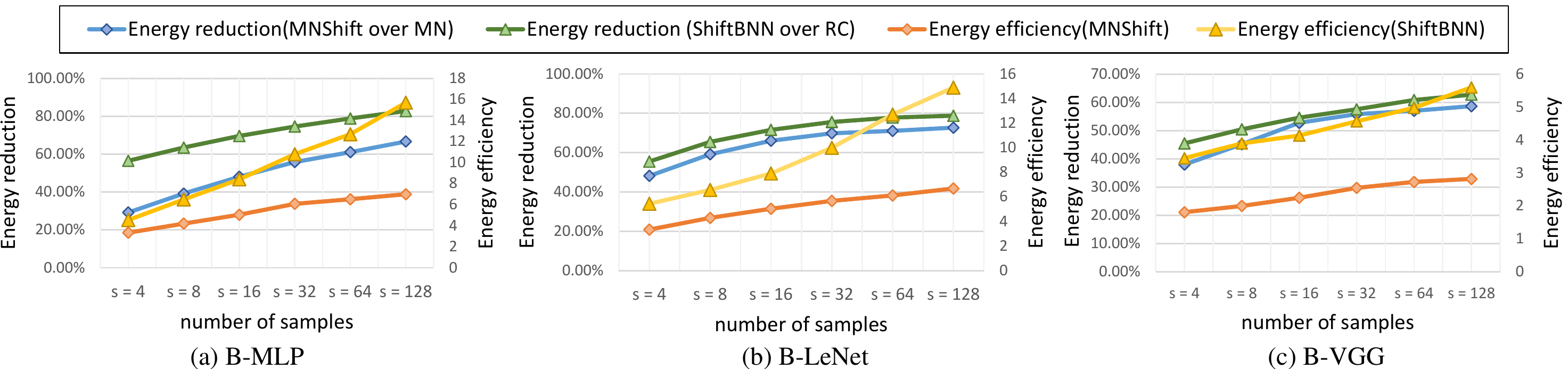}
\vspace*{-3.5mm}
\caption{The energy reduction of Shift-BNN (MNShift-Acc) over RC-Acc (MN-Acc) and energy efficiency of Shift-BNN and MNShift-Acc when training with different sample size.}
\label{fig:scalar}
\vspace*{-4.5mm}
\end{figure*}

\begin{figure}[t]
\centering
\includegraphics[scale=0.6]{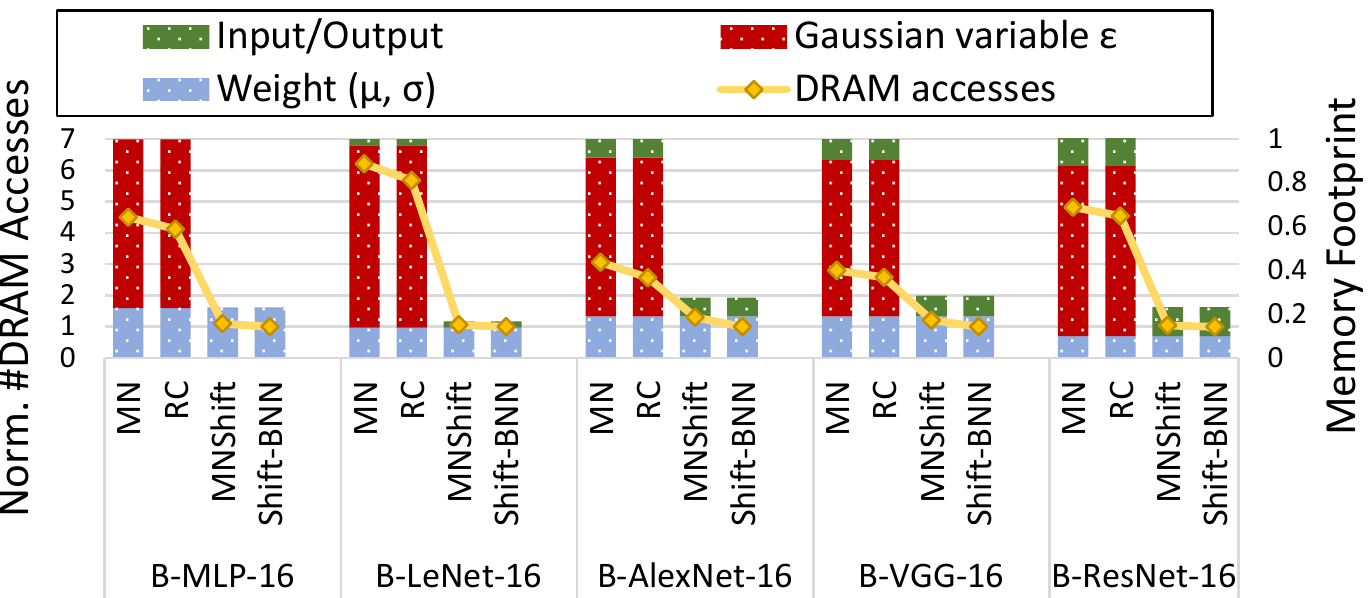}
\vspace*{-3.5mm}
\caption{The effectiveness of our LFSR reversion strategy on reducing DRAM accesses and memory footprint.} 
\label{fig:fp}
\vspace*{-6.5mm}
\end{figure}

%\noindent\textit{1) Effectiveness on energy consumption and performance}
\textbf{\underline{Effectiveness on energy and performance}.}
Fig. \ref{fig:energy} illustrates the energy consumption of Shift-BNN compared against other accelerators. As it shows, the Shift-BNN accelerator achieves an averagely 62\% (up to 76\%), 70\% (up to 82\%), and 39\% (up to 44\%) energy consumption reduction compared with RC-accelerator (RC-Acc), MN-accelerator (MN-Acc), and  MN-Shift-accelerator (MNShift-Acc), respectively. The outstanding energy reduction of Shift-BNN is from the elimination of $\epsilon$'s DRAM accesses by using our LFSR reversion strategy. %Furthermore, compared with MN-accelerator (MN-Acc), Shift-BNN achieves an averagely 71\% (up to 82\%) energy savings. Secondly, the Shift-BNN accelerator reduces 38\% (up to 44\%) energy consumption compared with MNShift-accelerator (MNShift-Acc) which also adopts our LFSR reversion technique. The reasons are two-folded. 
 The MNShift-Acc reduces the energy consumption by 53\% averagely compared with MN-Acc which is less than that of Shift-BNN over RC-Acc (i.e., 62\% reduction). This implies that our LFSR reversion technique is also effective on MN-accelerator but reaps less energy saving than applying on RC-accelerator. As discussed in Section \ref{archi}, this is caused by the large design overhead, e.g., duplicated adder trees, etc, when applying LFSR reversion strategy to MN-mapping scheme. We further observe that Shift-BNN achieves 68\% and 70\% energy consumption reduction over RC-Acc when evaluating on B-MLP and B-LeNet models, respectively. These number are larger than that of other BNN models. This is because $\epsilon$ takes a larger portion in the total off-chip data transfer, and off-chip memory access consumes a larger portion of total training energy consumption for B-MLP and B-LeNet. 
%  Lastly, we note that RC-Acc produces lower energy consumption than the baseline MN-Acc by 22\% averagely thanks to its better data reuse, except in fully connected B-MLP models where there is no opportunities for input neuron reuse. 
 
Since Shift-BNN mainly targets on reducing the data transfer during training, it is interesting to see if the data transfer reduction can be converted to performance improvement. Fig. \ref{fig:speedup} shows the speedup of Shift-BNN over other accelerators. From the figure, we observe that Shift-BNN accelerator achieves an average 1.6$\times$ (up to 2.8$\times$) speedup over RC-Acc. We found that the reduced execution time mainly comes from the removal of all memory accesses of $\epsilon$ in FC layers. As we know, the memory access of $\epsilon$ and the computation in a certain layer can be done in parallel by using double-buffering. Thus, in the computation-dominated convolutional layers, removing the memory access of $\epsilon$ may not reduce the latency. However, in the parameter-dominated FC layers, the memory access (including storing in FW and fetching in BW) time of \textit{S samples} of $\epsilon$ significantly exceeds the computation time since the number of MACs in FC layers are much smaller than that of convolutional layers. For example, the memory access time of $\epsilon$ is 8 $\times$ over computation time in the 1st layer of B-MLP-8. Accordingly, there is an obvious variance in the performance improvement across different BNN models. For instance, for the fully-connected B-MLP models, the Shift-BNN gains the maximum 2.6$\times$ speedup on average, while for the convolution dominated B-VGG and B-ResNet models, Shift-BNN achieves an averagely 1.18$\times$ performance improvement.
% We further observe that MNShift-Acc achieves almost the same amount of performance improvement over MN-Acc as that of Shift-BNN over RC-Acc (i.e., 1.7$\times$), which indicates the speedup achieved by our proposed method will not be affected by different dataflow or a more complex architecture design. However, since the average performance of MN-Acc is inferior to RC-Acc due to the low resource utilization, Shift-BNN still achieves averagely 1.6$\times$ speedup over MNShift-Acc, which is actually the performance gap between RC-Acc and MN-Acc. 

Fig. \ref{fig:ee} shows the energy efficiency of Shift-BNN accelerator compared with other designs. The energy efficiency is defined as throughput per watt (GOPS/Watt). It is shown that Shift-BNN boosts the energy efficiency by 4.9$\times$ (up to 10.8$\times$), 10.3$\times$ (up to 26.1$\times$) and 2.5$\times$ (up to 4.6$\times$) averagely compared with RC-Acc, MN-Acc and MNShift-Acc, respectively. The highest energy efficiency achieved by Shift-BNN is observed on B-MLP-32 model, which enjoys significant reductions on both energy consumption and latency. Furthermore, Shift-BNN also yields averagely 4.7$\times$ energy efficiency compared with Telsa P100. We observe that the GPU outperforms the baseline when training deeper BNNs with larger sample size because of its highly parallel computing and sufficient memory bandwidth. However, it is still beaten by the variants of Shift-BNN that are equipped with our techniques, e.g., even MNShift outperforms GPU by 1.9$\times$ energy efficiency. This is because the off-chip memory access of a large amount of GRVs can not be avoided when training BNNs on GPUs either.  

\textbf{\underline{Reduction of DRAM accesses and memory footprint}.}\\
Fig. \ref{fig:fp} shows the number of DRAM accesses and memory footprint breakdown of four accelerators when training on BNN models with 16 samples. For the DRAM accesses, we observe that the MN-Acc and RC-Acc always require much more DRAM accesses than Shift-BNN and MNShift-Acc in different BNN models. For example, the number of DRAM accesses in MN-Acc (RC-Acc) are 5.7$\times$ (5.8$\times$) larger than that in MNShift-Acc (Shift-BNN) in the $\epsilon$-dominated B-LeNet-16 model. Even in the wider and deeper models (e.g., B-VGG-16 and B-ResNet-16) where the weight parameters and intermediate feature maps occupy a considerate portion of total data transfer, Shift-BNN still gains averagely 2.6$\times$ reduction on DRAM accesses.
The significant reduction of DRAM access is the major source of Shift-BNN's high energy efficiency. As various lower-precision training techniques \cite{banner2018scalable,yang2020training, fu2020fractrain}, e.g., 8-bit integer training, have been proposed recently, the cost of MACs could become much less. Thus the memory saving techniques of Shift-BNN could have more benefits once these techniques are extended to BNN models.
Furthermore, as the figure shows, both Shift-BNN and MNShift-Acc reduce averagely 76.1\% memory footprint during training compared with accelerators without LFSR reversion technique. 
From the figure, we can observe that the memory footprint taken by Gaussian variable $\epsilon$ is completely eliminated by MNShift-Acc and Shift-BNN.

\textbf{\underline{Scalability to larger sample size}.} \label{scalability}
In some high-risk applications, one may need a more robust BNN model to make decisions, thus requires training BNNs with a larger sample size to strictly approximate the loss function in Eq.\ref{eq2}. We evaluate three BNN models including B-MLP, B-LeNet and B-VGG by training them with different number of samples and report the corresponding energy consumption reduction and energy efficiency under different hardware designs. As can be seen, for all three models, the energy reduction achieved by both MNShift-Acc and Shift-BNN increases as the sample size becomes larger. For example, the energy savings increase from 55.5\% to 78.8\% as the sample size grows from 4 to 128 in B-LeNet. The outstanding scalability of our LFSR reversion technique is because of the increasing ratio of $\epsilon$ in the total off-chip memory accesses when we use more samples. We observe the similar increase in energy efficiency for MNShift-Acc and Shift-BNN as the training sample increases. Lastly, compared with MNShift-Acc, Shift-BNN achieves higher energy efficiency with various sample sizes. 

\begin{table}[h]
\vspace*{-4.5mm}
\caption{Resource usage of Shift-BNN components.}
  \vspace*{-1.5mm}
  \scriptsize
  \centering
  \begin{tabular}{|c|c|c|c|c|c|}
    \hline
    \textbf{Resource} & \textbf{PE} & \textbf{Shift} & \textbf{Function} & \textbf{GRNGs} & \textbf{NBin} \\
    \textbf{} & \textbf{tile} & \textbf{array} & \textbf{units} &  &  \textbf{/NBout}\\
    \hline
    \hline
    LUT & 966 &	222	& 785 & 2277 &	0  \\
    \hline
    FF & 469 & 464 & 399 & 4224 & 0	\\
    \hline
    DSP & 16 & 0 & 32 & 0 & 0  \\
    \hline
    BRAM & 0 & 0 & 0 & 0 & 48 \\
    \hline
    $\mathbf{P_{avg}}$ (W) & 0.076	& 0.016	& 0.008	& 0.005	& 0.112 \\
    \hline

  \end{tabular}
  \vspace*{-2.5mm}
  \label{table:resource}
\end{table}

\textbf{\underline{Resource usage and power}.}
Table \ref{table:resource} lists the resource usage and average power of different hardware modules in one SPU. As can be seen, the shift units array and function units consume less LUT and FF resources compared with the PE tile. Although the function units requires more DSPs to implement function units due to the sampling, derivative calculation and updating processes, their average power dissipation is much smaller than that of PEs since only 1 of 16 function units is activated during convolutional layers. The similar effect can be observed on GRNGs whose average power is only 0.005W, albeit occupying more LUT and FF resources than others.

\vspace*{-1.5mm}
\section{Related works}
\textbf{Accelerators for BNNs.}
There is an increasing demand for designing specific BNN accelerators recently. VIBNN \cite{cai2018vibnn} optimizes the hardware design of GRNGs and proposes an FPGA-based implementation for BNN inference. FastBCNN \cite{wan2020fast} targets on accelerating the BNN inference via neuron skipping technique. \cite{yang2020all} proposes a BNN inference accelerator by leveraging post-CMOS technology. Different from the above efforts, our work proposes a highly efficient BNN accelerator that focuses on optimizing the training procedure.

\textbf{DNN training optimization} has been extensively studied \cite{song2019hypar,qin2020sigma,zhang2019eager,yang2020procrustes,mahmoud2020tensordash}. For example, eager pruning \cite{zhang2019eager} and Procrustes \cite{yang2020procrustes} exploit the weight sparsity during the training stage by leveraging aggressive pruning algorithms and develop customized hardware to improve the performance. Procrustes also employs LFSR-based GRNGs but in purpose of enabling weight initialization and decay.
% TensorDash \cite{mahmoud2020tensordash} further exploits the sparsity in weights, activations and gradients during training and designs an efficient hardware scheduler to handle the computation irregularity.
Since our work reveals the key challenge in BNN training and mainly focuses the reducing the data transfer of $\epsilon$ which is irrelevant to sparsity, the above works are orthogonal to ours. %In contrast, our work reveals the key challenges in BNN training, i.e., the huge data transfer of $\epsilon$, and eliminates all of it to achieve highly efficient BNN training.

\textbf{Reducing DRAM energy consumption}
Many works focus on addressing costly DRAM accesses during the DNN inference or training process. 
% vDNN \cite{rhu2016vdnn} targets on handling the large amount of intermediate data that need to be sent to DRAM when training in batches on GPU, while in BNNs $\epsilon$ dominates the off-chip transfer due to the sampling processing. 
EDEN \cite{koppula2019eden} leverages approximating DRAM technique to reduce the energy and latency while strictly meets the target accuracy.
Shapeshifter \cite{lascorz2019shapeshifter} explores the opportunities in shortening the transferred data width during DNN inference. These works are orthogonal to ours since we explore the unique feature of BNN training and eliminate intensive data transfer without accuracy loss from a different perspective. 

\vspace*{-2mm}
\section{Conclusion}
In this paper, we reveal that the massive data movement of GRVs is the key bottleneck that induces the BNN training inefficiency. We propose an innovative method that eliminates all the off-chip memory accesses related to the GRVs without affecting the training accuracy. We further explore the hardware design space and propose a low-cost and scalable BNN accelerator to conduct highly efficient BNN training. Our experimental results show that our design achieves averagely 4.9$\times$ (up to 10.8$\times$) boost in energy efficiency and 1.6$\times$ (up to 2.8$\times$) speedup compared with the baseline accelerator.

\begin{acks}
This research is partially supported by NSF grants CCF-2130688, CCF-1900904, CNS-2107057, University of Sydney faculty startup funding, and Australia Research Council (ARC) Discovery Project DP210101984.
\end{acks}

\bibliographystyle{ACM-Reference-Format}
\bibliography{ref}

% \appendix









\end{document}